\documentclass[a4paper,12pt]{article} 
\usepackage{amsfonts,slashed}
\usepackage{url}
\usepackage{latexsym}
\usepackage{amsmath}
\usepackage{amssymb}
\usepackage{indentfirst}
\usepackage{graphicx}
\usepackage{epsfig}
\usepackage{amsthm}
\usepackage{amsfonts}
\usepackage{indentfirst}
\usepackage[colorlinks]{hyperref}
\usepackage{cite}
\usepackage{cancel}
\usepackage{tikz}
\usepackage{enumitem}

\usepackage{color}
\usepackage{slashed}

\usepackage{amsfonts,slashed}
\usepackage{url}
\usepackage{latexsym}
\usepackage{amsmath}
\usepackage{amssymb}
\usepackage{indentfirst}
\usepackage{graphicx}
\usepackage{epsfig}
\usepackage{amsthm}
\usepackage{amsfonts}
\usepackage{indentfirst}
\usepackage[colorlinks]{hyperref}
\usepackage{cite}
\usepackage{cancel}
\usepackage{tikz}
\usepackage{enumitem}

\usepackage{color}
\usepackage{slashed}

\numberwithin{equation}{section}

\newcommand{\commas}{“}

\def\ii{{\rm i}}

\begin{document}

\begin{flushright}
\today
\\
REVISED
\end{flushright}

\vskip 10mm
\begin{center}
{\Large\textbf{   {Supergravity in the Geometric Approach}}}
\vskip 5mm
{\Large\textbf{  and its Hidden Graded Lie Algebra}}
\vskip 1 cm
{L.\ Andrianopoli}$^{[a,b]}$,
{R. D'Auria}$^{[a]}$
\end{center}

\vskip 0.3in

\noindent
{\small
$^{[a]}$ {Politecnico di Torino, Corso Duca degli Abruzzi 24, 10129 Torino, Italy}\\
$^{[b]}$ {INFN, Sezione di Torino, Via P. Giuria 1, 10125 Torino, Italy}
}
\vskip 1in
\begin{center}
{\bf{Abstract}}
\end{center}
 In this contribution, we present the geometric approach to supergravity. In the first part, we discuss in some detail the peculiarities of the approach and apply the formalism to the case of pure supergravity in four space-time dimensions.
In the second part, we extend the discussion to theories in higher dimensions, which include antisymmetric tensors of degree higher than one, focussing on the case of eleven dimensional space-time. Here, we report the formulation first introduced by R. D'Auria and P. Fr\'e in 1981, corresponding to a generalization of a Chevalley-Eilenberg Lie algebra, together with some more recent results, pointing out the relation of the formalism with the mathematical framework of $L_\infty$ algebras.

\vfill
\noindent
{\footnotesize{\tt laura.andrianopoli@polito.it};\\
{\tt riccardo.dauria@polito.it}; }

\numberwithin{equation}{section}
\clearpage

\section{Introduction}
\par It is more than half a century since  {superstring theory \cite{Green:1987sp}, together with its strictly related low energy description, supergravity \cite{Freedman:1976xh,Deser:1976eh},  appeared and soon imposed themselves as some of the most investigated fields of research in High-Energy Physics}. Many important results  {at the frontier between Physics and Mathematics} have been obtained in the years in this field.

The present contribution will concern the construction of supergravity theories through the use of geometric concepts and structures  only \cite{Neeman:1978njh,{Castellani:1991et}}. The approach was first introduced in $D=4$ space-time dimensions as a  Group Manifold Approach, where the structure group of the theory is a \emph{graded Lie group} (a Lie supergroup)\footnote{In the following, we will adopt either the physical suffix \emph{super}- or the mathematical suffix \emph{graded}- interchangeably. }. 
However, when the supergravity theory is built in a higher dimensional space-time \cite{Cremmer:1978km}, the formalism has to be generalized to what is known in Physics literature as FDA approach \cite{D'Auria:1982nx} \footnote{
\label{fda}FDA is an abbreviation of \emph{Free differential algebra}. Strictly speaking, the name could be misleading, as it is a \emph{differential-graded algebra } (dg-algebra) which  in general is free only as a graded-supercommutative superalgebra, not as a differential algebra.
Several years after its introduction in the supergravity context, in \cite{D'Auria:1982nx}, this structure was recognized to be equivalent to a mathematical structure called  $L_{\infty}$ algebra.
The relation between super
$L_\infty$-algebras and the “FDA”s of the supergravity literature was made explicit in \cite{Fiorenza:2013nha}. We will elaborate further on this in the second part of the present contribution.
Here,  we will keep the  name  “FDA"  to be easily understood in the supergravity community.} since it is based on a higher algebraic structure than ordinary Lie algebras, so that an underlying structure group cannot be defined \commas a priori" but, as we will discuss in the last Section, can be recovered, \commas a posteriori", in a larger sense. What is peculiar in our approach to supergravity,   is its \emph{geometric} flavour from a mathematical point of view.

\subsection{Some History}
We recall that there have been actually several approaches to the construction of Supergravity theories: The well-honored Noether method was applied
to construct the first instance of a supergravity theory in \cite{{Freedman:1976xh},{Deser:1976eh}};  {then} the so-called Superspace approach  {appeared} \cite{Sohnius:1985qm},\cite{Wess:1992cp}, which features the use of an enlarged space parameterized by Grassmann odd coordinates $\theta^\alpha$ together with the usual space-time coordinates $x^\mu$; and the superconformal approach of the Belgian-Dutch school \cite{Freedman:2012zz}, where the superspace theories are obtained by gauge-fixing models enjoying a larger, superconformal invariance.

 Last, in order of time, is the so-called \emph{Geometric or Rheonomic  approach}.
 It was proposed in the year 1978  by Y. Ne'eman and {T.} Regge \cite{Neeman:1978njh}, as a new approach to the formulation of gauge
theories acting non-trivially on space-time, specifically gravity and supergravity. It is based {on the} formalism introduced by E. Cartan \cite{Cartan:1923zea} for the formulation  of Riemannian geometry in a completely geometrical setting. Cartan's approach implies a geometrical and group-theoretical way of formulating  General Relativity. Indeed, as the adopted formalism relies on the use of \emph{ differential forms}, Cartan's beautiful setting is independent of  {the choice of} a given coordinate frame. At the same time, it gives a prominent role to the gauge invariance of the theory under the Lorentz group, which emerges quite naturally from the formalism. As a matter of fact, in Cartan's view, Riemannian geometry has to be seen as pertaining to finite dimensional Lie groups rather than to the infinite dimensional group of general coordinate transformations (GCTG in the following). In the latter case, it would be difficult to see how gravitation could be unified with gauge theories of other interactions, at least at the classical level, what instead seems quite natural in the geometrical formalism developed by Cartan.
\par
Following this line of approach{,} Y. Ne'eman and {T.} Regge further developed Cartan's formalism proposing that {it should be possible in principle to construct} any diffeomorphic and gauge invariant theory directly on a \emph{ group manifold} $\rm G$, the physical fields being defined as the Lie algebra valued gauge fields in the coadjoint representation {of the} group. Therefore  their original formulation  was denoted   \emph{Group Manifold Approach}. The above geometric formalism was then further developed in \cite{DAuria:1980cmy}, where the role of the graded Lie algebra cohomology for the construction of supergravity theories was put in evidence.

Coming back from IAS to Torino University in 1978, Tullio Regge proposed to one of the authors ({R D'A.}) to develop the approach extensively, namely in any space-time dimension $4<D\leq 11$ with any number allowed of supersymmetry generators \footnote{The restriction to 11 space-time dimensions is due to the fact that, for $D > 11$, supersymmetric theories necessarily include fields with helicity higher than 2.}  and in the presence of matter sources.  His legacy was then further developed by his research group  in the Physics Department of Torino University (mainly by R.D'A and P. Fr\'e), and later by the Torino-Politecnico group.

Using the geometric formalism, it was possible indeed to also rewrite the pure \footnote{By pure (super)-gravity we mean a theory with no matter couplings.} supergravity theories in  five \cite{Cremmer:1980gs} space-time dimensions in a simple and elegant geometric way \cite{DAuria:1981yvr}, based on the Maurer-Cartan equations satisfied, in the vacuum, by the 1-form fields dual to the generators of the structure group. Since then, the systematic use of the geometric and group{-}theoretical approach has been an essential tool to obtain many interesting results in supergravity. Most of the supergravity theories in every dimensions $D\leq 11$ (see \cite{Salam:1989fm} for a comprehensive review of the first achievements in supergravity) were  reformulated or constructed from scratch  within the geometric approach. Some of them are collected in \cite{Castellani:1991et}. Often, the use of the geometrical approach  allowed  to give a complete {answer} to problems where other approaches  had given only limited answers.  This was particularly fruitful when matter coupled supergravity theories were considered, in which case the geometric approach allowed to put in light all the global and local  non-linear symmetries governing their  inte\-raction.
\cite{{Ceresole:1992su},{Ceresole:1993qq},{Ceresole:1993nz},{Ceresole:1995jg},{Billo:1995ge},{Ceresole:1995ca},{Cadavid:1995bk},{Andrianopoli:1996ve}}

A typical example was the  construction of the ${\mathcal{N}}=2$, $D=4$ matter-coupled supergravity \cite{DAuria:1990qxt,Castellani:1991et,Andrianopoli:1996cm} which was previously formulated using the superconformal approach in a coordinate dependent way \cite{deWit:1984rvr}. The geometrical approach provided a complete Lagrangian (including all the fermions contributions) and the transformation laws leaving it  invariant under supersymmetry, quite independently of the coordinates used, not only referring to the space-time frame, but also  to the  scalar fields description, which in these theories is generally
associated with a non-linear
$\sigma$ model,  with specific geometric features.
Within the geometric approach, it was very natural to find, among the conditions for supersymmetry invariance of the theory, a set of differential and algebraic relations fully characterizing the scalar $\sigma$-models of the matter-coupled theory: regarding the scalars in the  vector multiplets, these correspond to the notion of  K\"ahler Special  Geometry while,  regarding those in the hypermultiplets, they were instead  recognized as the defining relations of Quaternionic manifolds. (As comprehensive reviews of the subject from a physicist's perspective, we refer to  reference \cite{Castellani:1991et}  and, for more recent results, to the excellent review  \cite{Trigiante:2016mnt}).\\

As supergravity theories, besides their being field theories \commas per se", are also the low energy limit of superstring theory, the results found about the scalar manifolds of supergravity theories  also give insight and have a counterpart description in terms of  Calabi{-}Yau compactifications of superstring's  target-space description.

Another interesting point is the question of whether the geometric approach is completely equivalent to the purely space-time approach. This seems not to be the case in some \emph{chiral} theories, like $\mathcal{N}=1$, $D=6$ {\cite{DAuria:1983jkr}}, and $D=10$, $IIB$ \cite{Castellani:1993ye}.
 What is common to these theories is their non-standard description, in terms of  Hodge-duality frame, of the gauge fields involved: the geometric approach, whose frame-independence can be extended also to the electric/magnetic duality frame, allowed to obtain new results, not accessible within other approaches:
As an example, in the pure, minimal $D=6$  supergravity, the gravity multiplet contains the sechsbein{,} a Weyl gravitino, and a 2-form potential (that is an antisymmetric two-index tensor) with a \emph{self-dual} 3-form field strength. Using the geometric approach in superspace, it was shown {\cite{DAuria:1983jkr}}
that the \emph{self-duality} of the 3-form field-strength, necessary to match the number of Bose-Fermi on-shell degrees of freedom, follows from the  variational equations in \emph{superspace},
but not from their space-time restriction. As a consequence, the theory is consistent in superspace, although its  {Lagrangian} restricted to space-time is not supersymmetric invariant off-shell. Exactly in the same way can be treated the $D=10$, $IIB$ theory  \cite{Castellani:1993ye}, so that also in this case the self-duality of the 5-form can be retrieved from the superspace equations of motion.

We stress that, as we are going to discuss in the following, the group manifold approach is a \emph{superspace} approach but, differently from  {other} superspace approaches, the (super)-fields entering the theory $\mu^A(x^\mu,\theta^\alpha)$ are never expanded in the Grassmann-odd coordinates $\theta^\alpha$
and no Berezin integration is necessary.

\vskip 5mm

However, the very real impact of the approach was realised, beginning of 1981, with the extension of the geometric method to supergravity in dimensions $D$ higher than five, namely $5<D\leq 11$ \cite{D'Auria:1982nx}. Indeed,
in the general case, the spectrum of supergravity theories includes  $p$-form potentials, of rank $p$ higher than 1, associated with graded-antisymmetric tensors
$$A^{(p)}= \frac 1{p!} A_{\mu_1\dots \mu_p}dx^{\mu_1}\wedge dx^{\mu_p}\,.$$
Their presence in the supergravity spectrum  makes the direct construction of the Lagrangian in terms of the Maurer-Cartan 1-forms  of a Lie (super)-algebra  (see Appendix \ref{mc}) problematic.
Indeed, when  higher  \emph{p-forms}, with $p>1$, are present in the physical spectrum,  we cannot  consider these fields as spanning the cotangent space  of a (super)-group manifold and therefore the construction based on the group manifold must be modified.

In the paper {\cite{D'Auria:1982nx} the authors devised how to overcome this obstacle by defining a new structure, generalizing to higher forms the Maurer-Cartan framework, in such a way as to include also $p$-forms, of any degree $p\geq 1$, in the set of   forms spanning the Maurer-Cartan set, thus generalizing this notion. The procedure to introduce the
  higher $p$-forms was obtained by inspecting a cochain system based on  an ordinary (graded) Lie algebra, following the results of the generalized  Chevalley-Eilenberg cohomology group of graded Lie algebras \cite{D'Auria:1982nx}. Furthermore, mimicking the Maurer-Cartan equations of an ordinary $(super)$ Lie algebra, they considered the \emph{exterior differential}, \commas $d$", of any $p$-form potential and required  it to be expressed as a polynomial in terms of the wedge product of all possible forms in the enlarged Maurer-Cartan set, consistently with their degree. Integrability is then obtained by  the cohomological requirement $d^2=0 $, thus providing a generalization of the dual form of the Jacobi identity.

 The structure so obtained  was given the name of \emph{Cartan integrable system} (CIS) and later \emph{free differential algebra} (FDA) (see footnote \ref{fda}). Soon after, this new formalism was also applied in \cite{vanNieuwenhuizen:1982zf},\cite{Castellani:1982kd}.

  Working on the FDA,  they were also able to show that the higher $p$-forms can be further expressed as polynomials in terms of 1-forms, thus reconstructing, from the eleven-dimensional super-Poincar\'e algebra, an extended Lie algebra which can be considered as the true Lie algebra of the $D=11$ supergravity.

The approach outlined above was undertaken in reference
\cite{D'Auria:1982nx},
where it was applied to formulate the eleven-dimensional, maximal theory of supergravity in superspace. This theory, earlier constructed on space-time  \cite{Cremmer:1978km}, was the first instance where an antisymmetric tensor field, here of rank three (a 3-form potential), appeared in the spectrum of the physical fields  as an essential ingredient to get a supersymmetry invariant theory. This is not a special feature of the eleven-dimensional theory only. An analogous treatment can be done to all supergravity theories where antisymmetric tensor fields appear in the spectrum, specifically to supergravity in space-time dimensions  $5<D\leq 11$, as it was  performed explicitly, for example, in the minimal theory in $D=7$ in \cite{Andrianopoli:2016osu}.

\vskip 5mm
Besides its applications in Physics, the approach  of \cite{D'Auria:1982nx} turned out to be interesting also from a mathematical point of view.
 Indeed, some years later, in the nineties,  a group of mathematicians realized \cite{Stasheff92},\cite{Lada:1992wc} that this kind of graded algebraic structure, the CIS  built in  \cite{D'Auria:1982nx}, being dated 1981, is actually \emph{the first historical example of an $L_\infty$ algebra} \cite{Sati:2008eg},\cite{Fiorenza:2013nha},\cite{nlab},\cite{CEalgebra}.
 The formalism used in\cite{D'Auria:1982nx} is actually dual to the standard formalism of $L_\infty$ algebras, which is given in terms of multi-brackets, since it is instead formulated in terms of a graded \emph{coalgebra} of differential $p$-forms, namely in terms of the space of $p$-{forms} dual to the generators of the $L_\infty$ algebra. As an example, in the original D=11 case studied in \cite{D'Auria:1982nx}, the algebra is constructed in terms of generators with form-degree three and six, besides the usual 1-form generators \footnote{ Note that this is in fact the formalism mostly used for the  formulation of the extended Chevalley-Eilenberg cohomology of the Lie algebras so that we may also say that the formalism is a generalization of the Chevalley-Eilenberg cohomology of Lie algebras.}.

 The mathematicians also pointed out  that the name of Free Differential Algebra (FDA), given in physical literature to the CIS structures, is not fully appropriate, since such structures are not \commas free" but only semi-free,  their underlying
 graded algebras being free. In the following, however, we shall not adopt the name of semi-free differential graded algebra (SFGDA), but
  for the sake of simplicity we will be faithful to the original name, well understood in the Physics community, and continue to call these structures FDA, the semifree character being understood.

\vskip 5mm
The rest of the paper is articulated as follows:
\\
In Section \ref{EC} we give a summary of the Einstein-Cartan formulation of general relativity, putting in evidence its geometric, group theoretical formulation. This will also set the stage for the extension of the formalism to supergravity, which is then the object of  Section \ref{Sugra}, where the formulation of supergravity in the geometric approach is presented.
In Section \ref{4}, we will extend the geometric formalism to supergravity theories including higher $p$-forms, where the theories are formulated as FDA's. In particular, we discuss in some detail  the case of D=11 supergravity and how it is determined by a FDA, thus giving for the first time an explicit formulation of  $L_\infty$ algebras.
We stress that the choice of the maximal theory in D=11 is motivated by the fact that supergravity in D=11 is not only the first theory where historically this new kind of structure has appeared, but even more, because D=11 supergravity is in a sense the most general supergravity theory.
Indeed, from this theory, by compactification of some of the spatial dimensions, one can obtain all the $D\leq 9$ dimensional supergravity theories.
Finally, in Section \ref{hidden} another important consequence of the approach
studied in \cite{D'Auria:1982nx} is reported, namely the possibility of trading a given  FDA into an equivalent ordinary graded Lie algebra. We collected in the Appendices our notations and conventions, together with some more technical details.

\section{Einstein-Cartan Gravity, a short Resum\'e} \label{EC}

In this Section we shall first remind some of the most important properties of the Cartan formulation of the Einstein gravity in order to establish the notations and thus setting the stage for the formulation of its extension to the Poincar\'e group manifold. This is a preparatory discussion in view of obtaining  the geometrical interpretation of supersymmetry (also called \emph{rheonomy}) in supergravity theories. We recall the principal properties of the standard Cartan-Einstein four-dimensional Lagrangian, which is the starting  point for our description, and of its extension to a theory defined on the full  Poincar\'e group.
Notations and conventions used here and in the following are given in Appendix \ref{notations}.

In the original Lagrangian formulation by Cartan,  the   field content is given by the spin connection and the vierbein, $\mu^A=\{\omega^a_{\,\,b},\,V^a\}$, which are 1-form fields
\begin{align}
    \mu^A\,:\quad \mathcal{M}_4 \to G\,,
\end{align}
where $\mathcal{M}_4$ is the four-dimensional space-time, and
$ G$ is the structure group, which in this case is the Poincar\'e group $G=\mathrm{ISO}(1,3)$. This means that we can identify the space-time manifold as the base space $\mathcal{M}_4$ of the principal  fiber bundle structure
$[{\mathcal M}_4, \mathrm{H}]$, whose fiber is the Lorentz group $ {H}=\mathrm {SO}(1,3)\subset G$. Here, the 1-form fields $\mu^A$ locally span the cotangent space to $G$.

The Poincar\'e group is generated by   the algebra $\mathfrak{iso}(1,3)$, with  Lorentz generators $J_{ab}$ and translation generators $P_a$,   satisfying
\begin{align}
    [J_{ab},J_{cd}]= -2\,\eta_{a[c}J_{d]b}+2 \,
    \eta_{b[c}J_{d]a}\,,\quad
[J_{ab},P_{c}]=-2\,P_{[a}\eta_{b]c}\,,\quad
[P_a,P_{b}]=0\,.\label{poin}
\end{align}

In the vacuum of the dynamical theory,  the 1-forms $\omega^{ab}, V^a$  span the cotangent space of $G$, so that:
\begin{align}
    \omega^{ab}(J_{cd})= 2\delta^{ab}_{cd}\,,\quad V^a(P_b) =\delta^a_b \,,
\end{align}
and they satisfy the Maurer-Cartan equations (see appendix \ref{mc}):
\begin{align}  d\omega^{ab} -\omega^a_c \wedge \,\omega^{cb}&=0\label{vaclordef}\\
 dV^a - \omega^a{}_b \wedge V^b&=0 \,.\label{vactordef}
\end{align}
This corresponds to their being left-invariant 1-forms on ISO(1,3).

Out of the vacuum, $\omega^{ab}, V^a$ are space-time valued 1-forms, corresponding to dynamical fields on space-time. They acquire  curvature so that they become \emph{non-left invariant} \footnote {Because of this, the  approach we are going to describe is sometimes named also \emph{(soft)} group manifold approach.}, their curvatures being the Einstein-Lorentz curvature 2-form $R^{ab}$ and the torsion 2-form $\mathring{T}^a$, defined as:
\begin{align} R^{ab}&\equiv d\omega^{ab} -\omega^a_c \wedge \,\omega^{cb}\label{lordef}\\
\mathring{T}^a&\equiv dV^a - \omega^a{}_b \wedge V^b= \mathcal{D}V^a\,,\label{tordef}
\end{align}
where we denoted by $\mathcal{D}V^a=d V^a-\omega^{ab}\wedge V_b$ the Lorentz-covariant differential of the vielbein. For a formal definition of the notions above, see Appendix \ref{mc}. A more detailed discussion of (non) left-invariant forms in gravity and supergravity  is given in Section \ref{Sugra}, while treating supergravity.

\vskip 3mm
The Einstein-Cartan Action is given, in terms of the above fields, by:
\begin{equation}\label{action}
    \mathcal {A}\,[\omega^{ab},V^a]= \frac{1}{4 \kappa^2}\int_{\mathcal{M}_4} R^{ab}\wedge V^c\wedge V^d {\epsilon}_{abcd}\,{,}
\end{equation}
where $\kappa=\sqrt{8\pi\,\mathcal{G}}$, $\mathcal{G}$ being the gravitational constant \footnote{In the following, we will often adopt natural units, where $\kappa= \hbar=c=1$.}. The integration is performed on the base space $\mathcal{M}_4$  of the principal bundle $[
\mathcal{M}_4,\,H]$, which is identified with the physical space-time.

Let us remind  some of the properties of the Einstein-Cartan {Lagrangian} 4-form, that is of the integrand of (\ref{action}):

\begin{itemize}
\item Being written in terms of differential forms, it is completely geometrical and therefore invariant under general coordinate transformations (space-time diffeomorphisms).
\item It is invariant under Lorentz gauge transformations, but \emph{non-invariant} under gauge translations.
\item Expanding the two form $R^{ab}$ along a basis of 2-forms on $\mathcal{M}_4$, that is:
\begin{align}
   R^{ab}=   R^{ab}{}_{cd} V^c\wedge V^d=    R^{ab}{}_{\mu\nu} dx^\mu \wedge dx^\nu \,,
\end{align}
one easily recovers the usual form of the Einstein-Hilbert Lagrangian. Indeed, we can then rewrite the Lagrangian 4-form in \eqref{action} as:
\begin{eqnarray}
	R^{ab} {\wedge} V^{c}\wedge V^{d} \epsilon_{abcd}&=&  R^{ab}_{\phantom{ab}ij}V^{i} V^{j} V^{c} V^{d}\epsilon_{abcd}=
 \nonumber \\
	 &=& R^{ab}_{\phantom{ab}ij}V^{i}_{\phantom{i}\mu} V^{j}_{\phantom{j}\nu}
	 V^{c}_{\phantom{j}\rho} V^{d}_{\phantom{j}\sigma}
	\,d^{4}x\, \epsilon^{\mu\nu\rho\sigma}\epsilon_{abcd}=%
	 \nonumber\\
	&=&-4 R^{ij}_{\phantom{ab}ij} \textrm{det}V d^{4}x
	 .\label{equiv}
\end{eqnarray}
If we denote  world-indices by Greek letters, we have
\begin{equation}
	R^{ij}_{\;\;ij}
	\equiv R^{\mu\nu}_{\;\;\mu\nu}
	= \mathcal{R} {,}
	\label{I.4.4}
\end{equation}
where $\mathcal{R}$ is the scalar curvature and $\textrm{det}(V)=\sqrt{-g}$ is the square root of the metric determinant ($g=\textrm{det}(g_{\mu\nu})$). Hence we get:
\begin{equation}	\int_{\mathcal{M}_{4}} R^{ab}\wedge V^{c}\wedge V^{d}\epsilon_{abcd} = -4 \int_{\mathcal{M}_{4}} \mathcal{R}\sqrt{-g}\,d x^{4}
	\, .
	\label{I.4.5}
\end{equation}
\end{itemize}

Let us now observe that the formal equivalence between the Cartan and Einstein-Hilbert formulations just shown does not mean that they are completely equivalent.

First of all, the Cartan  formalism in terms of the vierbein 1-form, exhibiting explicit gauge invariance  under Lorentz transformations, makes it possible to introduce spinors in the {G}eneral {R}elativity framework, contrary to what happens in the usual formalism. Indeed in the world-index setting, tensors transform under {$\rm {GL}(4,\mathbb{R})$}, while spinors are in a  representation of $Spin(4)\simeq \mathrm{SL}(2,\mathbb{C})$, which is the double covering of the Lorentz group $\rm SO(1,3)$ and therefore they can be naturally coupled in a formalism when Lorentz $\rm SO(1,3)$ covariance is present.

 Furthermore, the Einstein-Cartan Lagrangian is a \emph{first order {Lagrangian}}, that is the gauge fields $\omega^{ab},V^a$, being members of the same Adjoint multiplet of the Poincar\'e group, are off-shell independent, as it is natural in a {geometric Action} like (\ref{action}).
 By \emph{geometric}, we mean that
 it is built only in terms of differential forms, their exterior differentials, and wedge products of them.

 The Einstein-Cartan Action is then the formulation of gravity where the symmetry structure of the theory, that is Poincar\'e group, is fully manifest  and linearly realized.
  In this line of thought, we could look for the possible generalizations of the  {pure } gravity Lagrangian.
  This can be investigated with a scaling argument, referring to the physical scale dimensions of the fields appearing in the Action, and comparing then the scale dimension of the possible extra contributions with that of the Einstein-Hilbert term which, in natural units, scales as $[L^2]$. The length scale of the fields and curvatures can be immediately obtained from the Maurer-Cartan equations: {as $dx^\mu$} has scale [$L^1$ ], then  the vierbein and the  torsion 2-form \eqref{tordef} must scale as [$L^1$] in lenghts units, while the connection $\omega^{ab}$ and the Riemann curvature \eqref{lordef} must scale as [$L^0$].  We then see that the Lagrangian in \eqref{action} scales as [$L^2$] while products of curvatures $R^A=(R^{ab},\mathring{T}^a)$ would have a different scaling \footnote{In principle, a 4-form term like $\mathring{T}^a\wedge \mathring{T}_a$ would have the correct length scale. However, such term would have the opposite parity with respect to the Einstein-Cartan one. We will not investigate further this case.} and should therefore be omitted, unless we allow some dimensional constants to enter the {Lagrangian}.
  In fact, dimensional constants, such as mass terms, naturally  appear when gravity  theories are coupled  to matter.
For pure theories  described in terms of massless fields only,  pure gravity  being the simplest case, a dimensional constant  of dimensions mass squared is allowed, the cosmological constant $\Lambda\sim [L^{-2}]$. In the Einstein-Cartan approach, this term can be included by adding to the Einstein-Cartan Lagrangian 4-form the term $ \frac 13\,\Lambda\epsilon_{abcd} V^a\,V^b\,V^c\,V^d $. This gives rise to a Einstein {Lagrangian} with a cosmological term \footnote{This kind of extension, however, can be easily shown to be equivalent to starting with the group manifold of a (anti) de Sitter group instead of the Poincar\'e group and will not change anything in the  mechanisms we are going to discuss both for gravity  as for supergravity. Indeed we may note that the Poincar\'e group $\rm ISO(1,3)$ is an {In{o}n\"{u}}-Wigner contraction of the $\rm SO(2,3)$ group.}.

{Let us now write down the equations of motion  derived from the action (\ref{action}). Varying the action with respect to $\omega^{ab}$ and $V^d$ we find, respectively:}
{\begin{eqnarray}
 \label{deltaomega} \frac{\delta\mathcal{A}}{\delta \omega^{ab}}=0\,&:&\quad \mathring{T}^c \wedge V^d \epsilon_{abcd} = 0 , \\
  \label{deltavier} \frac{\delta\mathcal{A}}{\delta V^d}=0\,&:&\quad R^{ab}\wedge V^c \epsilon_{abcd} = 0 \,.
\end{eqnarray}
equation  \eqref{deltaomega}, after expansion  of the torsion 2-form along the vierbein:
\begin{align}
    \mathring{T}^c= \mathring{T}^c{}_{\ell m}V^\ell\wedge V^m\label{torEC}
\end{align}
reads:
\begin{align}
    \mathring{T}^c{}_{\ell m} V^\ell\wedge V^m\wedge V^d \epsilon_{abcd}=0
\end{align}
and,  writing the 3-form $ V^\ell\wedge V^m\wedge V^d$  as
\begin{equation}
\label{omega3}V^\ell\wedge V^m\wedge V^d=\epsilon^{\ell m d p}\,\Omega^{(3)}_p\,,
\end{equation}
where $\Omega^{(3)}_p$ is a three-dimensional hypersurface element of space-time,
gives:
\begin{equation}
\mathring{T}^c{}_{\ell m}
\epsilon^{\ell m d p}\epsilon_{abcd}\,\Omega^{(3)}_p=0\,,
\end{equation}
that is:
\begin{align}
    \left(\mathring{T}^p{}_{ab}+2 \mathring{T}^c{}_{c[a}\delta^p_{b]}
    \right)\Omega^{(3)}_p=0
\end{align}
whose solution, since $\Omega^{(3)}_p\neq 0$, is $T^a{}_{bc}=0$, that is $T^a=0$.
Then, the vanishing of the torsion, which  allows to write the spin connection 1-form in terms of the vierbein $V^a_\mu$ and its derivatives, is a consequence of the variational principle.

With an analogous computation, from equation \eqref{deltavier} one finds:
\begin{align}
   R^{ab}\wedge V^c \epsilon_{abcd}=0 \quad \Rightarrow \quad   R^{ab}{}_{\ell m} V^\ell \wedge V^m\wedge V^c \epsilon_{abcd}=0
\end{align}
that is, using again \eqref{omega3}:
\begin{align}
  - 6 R^{ab}{}_{\ell m}\delta^{\ell m p}_{abd}=0\,,
\end{align}
which can  be rewritten, in terms of the Ricci tensor $\mathcal{R}^a{}_b \equiv R^{ac}{}_{bc}$ and of the Ricci scalar $\mathcal{R}\equiv \mathcal{R}^a{}_a $, as:
\begin{align}
 \mathcal{R}^a{}_b -\frac 12 \delta^a_b \,\mathcal{R}=0   \label{Eeq}
\end{align}
that is like the Einstein equations in the absence of matter sources.

 It is important to stress that, besides the obvious diffeomorphism invariance, which is implicit since the Einstein-Cartan Lagrangian is coordinate-independent, being written in terms of differential forms,  the Lagrangian is invariant  under the fiber group $\rm{SO}(1,3)$, but not under the full Poincar\'e group, which is however the structure group,  all the fields being  valued in the  co-Adjoint representation of the Poincar\'e group. This
 follows from the fact that the Lagrangian includes the tensor $\epsilon_{abcd}$ which is  a Lorentz-invariant but not Poincar\'e-invariant tensor.
 This property can be easily checked  by considering an infinitesimal  Poincar\'e transformation on the gauge fields $\mu^A\equiv(\omega^{ab},V^a)$, where $A=([ab],a)=1,\cdots,10$ labels the co-Adjoint representation of the Poincar\'e group.
 {Defining} $\epsilon^A=(\epsilon^{ab},\epsilon^{a})${, being $\epsilon^{ab}$ and $\epsilon^{a}$} the parameters of the infinitesimal Lorentz and translation gauge transformations, respectively, we have:
\begin{equation}\label{adj1}
   \delta^{(gauge)} \mu^{{A}} = \left(\nabla \epsilon\right)^A {,}
\end{equation}
 where we denoted by   $\nabla$  the Poincar\'e gauge covariant differential. Decomposing the co-Adjoint index $A$ as indices of the Lorentz subgroup, from (\ref{adj1}) it follows
\begin{eqnarray}
 \delta^{(gauge)} \omega^{ab} &=& \mathcal D\epsilon^{ab} {,} \nonumber \\
\delta^{(gauge)} V^a &=& \mathcal D\epsilon^a + \epsilon^{ab} V_b \,,\nonumber
\end{eqnarray}
where, we recall, $\mathcal D= d-\omega$ denotes the Lorentz covariant differential.

It is then easy to see that the Lagrangian (\ref{action}) and the equations of motion are  invariant under  gauge  Lorentz transformations, but are not invariant under a gauge translation. Indeed, performing an infinitesimal gauge transformation on the Einstein-Cartan action \eqref{action},
 we have, up to total derivative:
\begin{equation}\label{trans}
 \delta^{(gauge)} \int R^{ab}\wedge V^c\wedge V^d  {\epsilon_{abcd}} =2 \int R^{ab}\wedge \mathcal D \epsilon^{c}\wedge V^d {\epsilon_{abcd}} =-2 \int \epsilon^c R^{ab} \wedge \mathring{T}^d {\epsilon_{abcd}} \neq 0\,,
\end{equation}
where we used  {the}
relation
$$\delta^{(gauge)} R^{ab}=\mathcal{D}^2 \epsilon^{ab}= 2R^{a}{}_c \epsilon^{cb}\,,$$
and we integrated by parts to get the last expression, which is not vanishing off-shell (we recall that the condition $\mathring{T}^a=0$ is found as an equation   of motion of \eqref{action}).

\section{D=4 supergravity in the geometric approach}\label{Sugra}

In this Section we will present the Ne'eman-Regge Group Manifold Approach to
minimal supergravity in $D=4$ space-time dimensions, introduced in \cite{Neeman:1978njh} and further elaborated and applied in \cite{Castellani:1991et}.

To this aim, we will first reconsider the Einstein-Cartan geometric description of gravity, discussed in Section \ref{EC}, pointing out that it could be reformulated as a theory where the 1-form fields are defined on the \commas soft-group manifold" $\tilde G$, locally equivalent to the structure group-manifold $G=\mathrm{ISO}(1,3)$.

Let us observe that, referring to the D=4 pure gravity theory, the full set of \commas generalized vierbeins" on G is given by the ten 1-forms
$\mu^A=(V^a,\omega^{ab})$. They span the cotangent space of $G$, whose directions can be parametrized by the coordinates $x^\mu$ ($\mu,\nu,...=0,1,2,3$ being general coordinate indices) associated to the action of the generators $P_a$, and by the coordinates $y^{\mu\nu}=-y^{\nu\mu}$ associated to the one of  the generators $J_{ab}$}.

A useful observation by Ne'eman and Regge is that the domain of the 1-form fields $\mu^A$ can be safely enlarged to be the full group manifold $G$. The consistency of this new point of view relies on the special form of the Einstein-Cartan Lagrangian, which is shared by its extension to supergravity:
As discussed in Section \ref{EC}, it is indeed a \emph{geometric}  Lagrangian defined on a principal fiber bundle whose fiber is the Lorentz group and, as such, the dependence on the Lorentz parameters is \emph{factorized}. The \emph{factorization} is reflected in the fact that the \emph{curvatures } $R^A$, which are a coadjoint multiplet of the structure group $G$,  can  in principle be expanded on a basis of 2-forms on the cotangent space to $G$:
$$R^A=R^A{}_{BC}\,\mu^B\wedge \mu^C=R^A{}_{bc}V^b\wedge V^c +R^A{}_{bc\,C}\,\omega^{bc}\wedge \mu^C $$
but
 their components in the directions of the Lorentz fiber are zero:
\begin{align}R^A{}_{ab \,C}=0\,,\label{facto}
\end{align}
so that they are fully described by their parametrization on space-time:
$$R^A=R^A{}_{ab} V^a\wedge V^b\,. $$
In an analogous way, supergravity can be constructed as an extension of gravity defined on the supergroup manifold $G=\overline{\mathrm{OSp} (1|4)}$  \footnote{The bar over $\mathrm {OSp}(1|4)$ means {In{o}n\"{u}}-Wigner contraction of the super Anti-de Sitter group $\mathrm{ OSp}(1|4)$ whose bosonic subgroup is $\mathrm{Sp}(4)\simeq \mathrm{SO}(2,3)$.},
whose graded algebra is given in terms of the generators $T_A=(J_{ab},P_a,Q_\alpha)$ as:
\begin{align}
& [J_{ab},J_{cd}]= -2\,\eta_{a[c}J_{d]b}+2 \,
    \eta_{b[c}J_{d]a}\,,\quad
[J_{ab},P_{c}]=-2\,P_{[a}\eta_{b]c}\,,\quad
[P_a,P_{b}]=0\,,\nonumber\\
&  [J_{ab},Q_\alpha]= \frac 12 (\gamma_{ab})_\alpha{}^\beta\, Q_\beta\,,\quad
[P_{a},Q_\alpha]=0\label{osp14}\\
&\{Q_\alpha,Q_\beta\}= - i (C\gamma^a)_{\alpha\beta}P_a\,.\nonumber
\end{align}
We have 10 bosonic and 4 fermionic tangent space directions in the supergrup $G$, and the cotangent space is spanned by the set of 1-forms $\mu^A=(\omega^{ab},V^a, \psi^\alpha)$
where $\psi^\alpha$ are the  1-forms dual to the generators $Q_\alpha$. In this case, the domain of the 1-form fields can be extended to be: $\mu^A=\mu^A(x^\mu,y^{\mu\nu},\theta^\alpha)$,  where $\theta^\alpha$ are the  Grassmann-odd parameters in the $\psi^\alpha$ directions. Therefore the 1-form fields $\mu^A=\{\omega^{ab}, V^a, \psi^\alpha\}$, and their curvatures $R^A$, can be thought of as superfields functions of the coordinates: $\mu^A= \mu^A(x^\mu, y^{\mu\nu};\theta^\alpha)$.
Note however that, as in the gravity case, the Lorentz group is factorized, that is it is on the fiber of a principal fiber-bundle structure.
As such, the curvatures can be expanded on a basis of 2-forms  in the \emph{physical} domain, which in the supergravity theories is named \emph{superspace} \footnote{In the minimal $D=4$ case  \emph{superspace} includes four Grassman-even and four Grassman-odd coordinates, and it will be denoted in the following as $\mathcal{M}_{4|4}$. In $N$-extended supersymmetric theories in $D=4$, the number of Grassman-odd coordinates is extended to $4N$, so that $\mathcal{M}_{4|4}\to \mathcal{M}_{4|4N}$.} and is spanned by the supervielbein $E^{\hat a}\equiv (V^a,\psi^\alpha)$ of the base space in the fiber bundle, with vanishing components in the $\omega^{ab}$ directions of $G$. We therefore have the following parametrization:
\begin{align}
    R^A(x,\theta)&= R^A{}_{\hat a \hat b}(x,\theta) E^{\hat a}\wedge E^{\hat b}\nonumber\\
    &= R^A{}_{(2|0) a  b} V^a \wedge V^b + R^A{}_{(1|1) a  \alpha} V^a \wedge \psi^\alpha +R^A{}_{(0|2)   \alpha\beta}\psi^\alpha \wedge \psi^\beta\,.
    \label{par}
\end{align}
Here, we denoted by  $ R^A_{(p,q)}$ the components of the curvature along $p$ bosonic vielbein $V^a$ and $q$ fermionic vielbein $\psi$.
In the following, we will name as \emph{inner} the components $R^A_{(2|0)ab}$, along the bosonic vielbein $V^a\wedge V^b$ only (and more generally, in higher dimensions, when higher forms are present, the components $R^A_{(p|0)a_1\dots a_p}$ with $p>2$  along $p$ bosonic vielbein only), naming instead as \emph{outer} the components along \emph{at least one} fermionic vielbein $\psi$, that is $R^A_{(1|1)a\alpha}$
 and $R^A_{(0|2)\alpha\beta}$.
  As we will clarify in the following subsection, the role of the inner and outer components of the superspace curvatures is not symmetric.
  What actually happens is that the outer  components of the {curvature} 2-forms turn out to be expressible algebraically, actually \emph{linearly}, in terms of the inner components of the set of curvatures.  As we will see in Section \ref{14action}, this is a consequence of the geometric structure of the Lagrangian and of its field equations.  This property is called \emph{rheonomy}, and will be further discussed in Section \ref{rheo}.
  For the sake of  brevity and simplicity we will show how this happens in  the simple example of pure ${\mathcal{N}}=1$, $D=4$ supergravity . However, the relevant results  hold exactly in the same way  for any supergravity theory, pure or matter coupled, in any dimension $ 4\leq D\leq 11$ and for any number  {$1\leq \mathcal{N} \leq 8$} of supersymmetry generators in the Lie {superalgebra}.

Let us now discuss how to formulate a general gravity or supergravity Lagrangian in the \emph{geometric} approach. It must respect  the following requirements:\\
\emph{The Lagrangian should be constructed using only wedge products of p-forms and  their exterior differential,  $d$, satisfying: $d^2=0$}.
Moreover, we require
\emph{the Hodge duality operator to be excluded from the construction of the Lagrangian}. This second requirement will be explained in a moment.
\footnote{Note that the Cartan-Einstein Lagrangian of Section \ref{EC}, and the $D=4$, $N=1$ Lagrangian that will be discussed in the present Section, both satisfy these requirements.}

The supergravity Action  in a superspace with $D$-space-time dimensions is  then obtained by integrating  the  Lagrangian   $D$-form on a $D$-dimensional bosonic hypersurface $\mathcal{M}_D$, immersed in superspace.
This in turn requires the introduction of appropriate \emph{embedding functions}, which should then be included in the set of fields in the Action integral, resulting in a theory containing extraneous fields devoid of any physical meaning.
However, using  a geometric Lagrangian in superspace, this problem is automatically overcome. Indeed, being geometric, the Lagrangian is invariant under diffeomorphisms in superspace \footnote{We often refer as \commas super-diffeomorphisms", in particular, to the diffeomorphisms along the odd directions of superspace.}, so that any variation of the embedding functions can be compensated by a diffeomorphism.
 This in turn implies, thinking of infinitesimal diffeomorphisms from a passive point of view, that \emph{any  surface of integration works equally well}.
Therefore, the equations of motion are valid in the full superspace, since, given any hypersurface, all other hypersurfaces in superspace can be reached by diffeomorphisms.
This clarifies why use of the Hodge operator should be avoided. Indeed, this condition follows from geometricity, because use of the Hodge duality operator implies choosing a given metric description (something that we would like to avoid in a geometric Lagrangian),  but also because it would make problematic the extension of the field domain from $\mathcal{M}_4$ to superspace.
 \footnote{ Actually, working with geometric lagrangians it is possible to formulate a \emph{generalized action principle} where the Lagrangian is integrated on a submanifold of the full group manifold $\rm G$.
Factorization of coordinates belonging to gauge subgroups of $\rm G$ can be proven to hold as the  equations of motion obtained from the variational principal imply that the curvatures are horizontal in the direction of the Lorentz generators, so that $\rm G$ actually becomes endowed with a fiber bundle structure. This horizontality property is not spoiled by the presence of supersymmetry.
For supergroups, where the notion of superspace as base space of the fiber bundle appears, the same procedure based on the extended action principle allows also to understand  why supersymmetry is not a gauge symmetry, since
it shows that the field-strengths (curvatures) in superspace are not horizontal in  the direction of the supersymmetry generators  of $\mathrm{G}$,  but can instead be expressed linearly in terms of the space-time components of the field-strengths, according to the principle of \emph{rheonomy} (see section \ref{rheo}). }

\vskip 3mm

In the following, we are going to construct explicitly the minimal, pure supergravity in $D=4$ space-time dimensions, with structure group   $G=\overline{\mathrm{ OSp}(1|4)}$, and principal fiber-bundle structure $[\mathcal{M}_{4|4},\mathrm{SO}(1,3)]$.

In this case the set of dynamical fields, defined in superspace $\mathcal{M}_{4|4}$, is given by the bosonic 1-forms $\omega^{ab}, V^a$, but also by the fermionic 1-form vielbein $\psi^\alpha$. They span the cotangent space of the structure group $G$, which in this case is the Super-Poincar\'e group. The set of curvatures $R^A$, which in the dynamical vacuum reduce  to the Maurer-Cartan equations of $G$ (see Appendix \ref{mc}), in this case turn out to be:
\begin{align}
    R^{ab}&\equiv d\omega^{ab} -\omega^a{}_c \wedge \omega^{cb}\hfill
 \nonumber   \\
 T^a&\equiv dV^a -\omega^a{}_b \wedge V^b  - \frac i2 \psi^\alpha (C\cdot\gamma^a)_{\alpha\beta}\wedge \psi^\beta= \mathcal{D}V^a - \frac i2 \bar\psi\gamma^a\wedge \psi\label{curv}\\
 \rho^\alpha&\equiv d\psi^\alpha -\frac 14 (\gamma_{ab})^\alpha{}_\beta\,\omega^{ab}\wedge \psi^\beta=\mathcal{D}\psi^\alpha\,,\nonumber
\end{align}
where $\mathcal{D}$ denotes the Lorentz covariant derivatives, acting differently on Lorentz vectors and Lorentz spinors. Here, the 1-form gravitino $\psi$ is a Majorana spinor, $\bar \psi\equiv \psi^t C$ being its adjoint spinor, $C$ is the charge-conjugation matrix and  $\gamma^a$ are the $\gamma$-matrices satisfying the Clifford algebra $\gamma_a\gamma_b +\gamma_b\gamma_a= 2\eta_{ab}$, see Appendix \ref{notations}. \footnote{To be clear, for example,  $\bar\psi\gamma^a \psi= \psi^\alpha (C\gamma^a)_{\alpha\beta}\psi^\beta$, where $C$ behaves as the metric of the spinor space, raising and lowering spinor indices.}
\footnote{We remark that the torsion supercurvature, that we name $T^a$,  differs from the torsion 2-form of gravity on space-time discussed in the previous Section,  since it contains a  bilinear current in the $\psi$ fields.
When discussing supergravity, to avoid confusion, we will reserve the symbol $T^a$ to the supertorsion in \eqref{curv},  referring to the purely bosonic torsion of the Riemannian geometry, that is  the Lorentz-covariant derivative of the vielbein,  as  $\mathring T^a \equiv \mathcal{D}\,V^a$.}

Consistency requires them  to satisfy the  following \emph{Bianchi identities in superspace}:
\begin{align}
&   \mathcal{D} R^{ab}=0
   \nonumber \\
&\mathcal{D} T^a+R^a{}_b \wedge V^b  - i \bar\psi  \gamma^a\wedge \rho=0 \label{bi1}\\
&\mathcal{D} \rho^\alpha+\frac 14 (\gamma_{ab})^\alpha{}_\beta\,R^{ab}\wedge \psi^\beta=0\,,\hfill\nonumber
\end{align}
 All the terms in the definition of the curvatures and in the Bianchi identities scale homogenously, since $\omega^{ab},V^a, \psi^\alpha$ and their curvatures have length scaling $[L^0],[L^1]$ and $[L^{1/2}]$, respectively.

We emphasize that, as it emerges from the above discussion, in the geometric setting only the Lorentz subgroup of the  (super-)Poincar\'e group turns out to be an actual gauge symmetry of the  theory, the spin connection $\omega^{ab}$ being its gauge connection. On the other hand, the vielbein 1-form $V^a$ and the gravitino 1-form $\psi^\alpha$ transform respectively as a vector and as a spinor under Lorentz transformations.
We will call this property   \emph{horizontality} condition.
This corresponds to the fact that, as stated in \eqref{par}, the curvatures $R^A$  have non-vanishing components not only  along the directions dual to the bosonic vielbein $V^a$, analogously to what we have seen in Section \ref{EC} (see in particular the discussion after equation \eqref{Eeq}), \emph{but also along the fermionic vielbein $\psi^\alpha$}.
This reflects a general property of supersymmetric Lagrangians, when realized in superspace: Supersymmetry invariance is not a \emph{gauge invariance} of the Lagrangian,  similarly to what happens with the translations. The Lagrangian indeed is not \emph{gauge}-invariant under the full algebra of commutators of $\overline{\mathrm{OSp}(1|4)}$, only the Lorentz subalgebra $\mathrm{SO}(1,3)$ being realized as a gauge symmetry.
As we are going to see, it is however invariant  under \emph{superdiffeomorphisms} in superspace.
This is analogous to what happens at the gravity level in the Einstein-Cartan formalism, where the theory is not invariant under \emph{gauge-translations}, but under \emph{diffeomorphisms} on space-time.

The new feature with respect to pure gravity is that the Lagrangian is invariant under local supersymmetry transformations. As we are going to show in the following, in the geometric approach the supersymmetry transformations on space-time are nothing but  \emph{diffeomorphisms in the odd directions of superspace}.
This means that the geometric approach allows for a geometrical interpretation of local supersymmetry, when formulated \emph{in superspace}, as a super-diffeomorphism, thus extending to the graded case the formulation of gravity as a theory invariant under space-time diffeomorphisms.
Just as the geometric Einstein-Cartan Lagrangian is not invariant under gauge translations but instead, being geometric,  is invariant under space-time diffeomorphisms, the supersymmetry invariance of the geometric Lagrangian in superspace has to be understood as an invariance under \emph{super-diffeomorphisms in superspace}, generated by the vector fields $\epsilon^\alpha D_\alpha$, and not as a gauge invariance.
To clarify this point,
 let us make here the following important distinction: We denote by $P_a, Q_\alpha$ the right-invariant (graded) generators dual to the left-invariant 1-forms of the translations and supertranslations respectively in the super-Poincar\'e group, while we denote by $D_a, D_\alpha$ their left-invariant counterpart, dual to the right-invariant 1-forms. \footnote{They satisfy the $\overline{\mathrm{OSp}(1|4)}$ superalgebra, with structure constants opposite to \eqref{osp14}.}
Both of them are invariant vector fields of the group manifold $G$ (that is, symmetries of the Maurer-Cartan equations satisfied by the fields  in the dynamical vacuum). However, due to the principal fiber bundle structure, out of the vacuum the vector fields $D_a, D_\alpha $ are not anymore invariant vectors of $G$.
They are instead   vector fields spanning the  tangent space of  superspace,  and dual to the  1-forms $V^a, \psi^\alpha$,
namely:
\begin{align}
    D_a(V^b)=\delta^b_a;\quad D_\alpha(\psi^\beta)=\delta^\beta_\alpha;\quad D_a(\psi^\alpha)=D_\alpha (V^a)=0\,. \label{rightgen}
\end{align}

Correspondingly, just as diffeomorphism transformations on the fields can be expressed by \emph{Lie derivatives} along $D_a$ directions, superdiffeomorpism transformations in superspace, that is supersymmetry transformations, can be expressed as \emph{Lie derivatives along odd directions $\epsilon\equiv\epsilon^\alpha D_\alpha$ of superspace}:
\begin{align}
    \delta_\epsilon \mu^A = \ell_\epsilon \mu^A\equiv d\left(\iota_\epsilon \mu^A
    \right)+\iota_\epsilon\left(d \mu^A
    \right)\,,\label{Lieder}
\end{align}
where we denoted by $\iota_\epsilon$ the contraction of a form along the odd tangent space direction $\epsilon$ so that, in particular:
\begin{align}
    \iota_\epsilon \psi^\alpha= \epsilon^\alpha\,,\quad \iota_\epsilon V^a=0\,.\label{iota}
\end{align}

An alternative definition of Lie derivative, which puts into light its differences with respect to gauge transformations, is given in Appendix \ref{Lie}.

\subsection{Supersymmetry as an on-shell symmetry}\label{on}

We could wonder if the geometric formulation of supergravity in superspace is fully equivalent to the standard formulation of supergravity on space-time, and in particular if supersymmetry transformations on space-time are completely equivalent to \emph{superdiffeomorphisms in superspace}. As we will see later in this Section, the answer is positive \emph{if we require, as already mentioned (in the paragraph below eq. \eqref{par}), that   the  parametrizations of the supercurvatures, equation \eqref{curv}, should be subject to the principle of \commas Rheonomy"}, whose meaning and use will be clarified later, in Section \ref{rheo}. Actually, Rheonomy is an intrinsic property of all the supergravity Lagrangians formulated in the geometric approach to superspace.

There is however an important difference between diffeomorphism invariance of a geometric theory on space-time and super-diffeomorphism invariance of a geometric theory in superspace:
In general, except in a few exceptional cases \footnote{The exceptional cases we are referring to are the off-shell supersymmetric theories, which close supersymmetry off-shell due to the presence of a set of auxiliary fields. These fields, when added to the coadjoint supermultiplet, make the supersymmetry transformations, leaving the Lagrangian invariant, to close the supersymmetry algebra \emph{off-shell} . This is related to the fact that the auxiliary fields allow to pair the number of off-shell degrees of freedom between boson and fermions. They are not dynamical degrees of freedom, as their equations of motion make them to vanish or to be expressed in terms of the physical fields. However, it does not {seem} possible to extend their introduction to theories with more than 8 supercharges nor to matter coupled supergravities (in particular when these include CPT  self-conjugate matter sources, unless we extend the superspace to an infinite number of extra bosonic directions. We will not discuss further off-shell supergravity in this review.
}, the supersymmetry algebra, when realized on dynamical fields,   is an \emph{on-shell} symmetry. This means that the closure of the exterior derivative operator in superspace: $d^2=0$, when applied on the defining fields of the theory, does not hold in general, but only on-shell, namely only if the equations of motion are satisfied. This corresponds, in the space-time description of the phenomenon, to the fact that the commutator of two supersymmetry transformations on the fields does not satisfy the Jacobi identities in general, but this property only holds  on shell, namely only if the equations of motion are satisfied.

This special feature reflects the peculiarity of supersymmetry, to be such that it maps into each other bosonic and fermionic degrees of freedom (d.o.f.), so that the supersymmetry representations, in general, should contain the same number of bosonic and fermionic degrees of freedom. These representations are called \emph{supermultiplets}, and collect several fields of different spin. This clashes with the fact that, in any Lagrangian theory, the number of d.o.f. of a given field is in general different if it is counted off-shell or after imposing its equations of motion: Just as an example, spinors halve their d.o.f. on-shell, while scalar fields do not change their d.o.f. at all.
Then, either we have off-shell matching of d.o.f., but then, in the general case, the classical trajectory cannot be supersymmetric, or we require on-shell matching. Since  we cannot give up the validity of the theory on-shell,\footnote{ Supergravity is an effective field theory, extending (classical) General Relativity.} the supergravity actions are constructed with the proviso that \emph{consistency of supersymmetry, that is closure of superdiffeomorphism invariance, only holds on-shell.}
This in turn has the consequence that the Bianchi identities \eqref{bi1}, and all their extensions to more general supergravity theories,
become identities only  after imposing the field equations.

As we will see, this special feature will turn out to be a resource of the supergravity theories, allowing to fully characterize all the properties of the classical theory even in the absence of a Lagrangian description.

 We stress again that these properties, that  will be explicitly shown in the following for the particular case of minimal $\mathrm{4D}$ pure supergravity, are shared by every supergravity theory in any possible number of dimensions and supersymmetry.

\subsection{N=1, D=4  Action in the  Geometric Approach.}\label{14action}

Let us now proceed here in finding the supergravity Action, and the supersymmetry transformation laws leaving it invariant.
The Action will be given by the integral, over a four dimensional bosonic submanifold of superspace, of a 4-form Lagrangian.

The starting point is the set of curvatures defined in equation \eqref{curv}, satisfying on-shell the Bianchi identities \eqref{bi1}.
To write down the {Lagrangian},  we require it to be \emph{geometric}.  Let us list here what it amounts to:
\begin{enumerate}
\item
It must be constructed using only differential forms, wedge product{s among them,} and the $d$ exterior differential;
\item
It must not contain the Hodge duality operator. This issue will be clarified in Section \ref{kin}.
\end{enumerate}
Other requirements of physical nature can be added which make easier, in more complicated cases, the search of the final form of the Lagrangian:
\begin{enumerate}
\setcounter{enumi}{2}
\item
First of all,
since the Einstein term, which must be always present,  in natural units {scales} as $[L^2]$, ({$[L^{D-2}]$ in $D$ dimensions}), all the terms in the Lagrangian must scale in the same way.
\item Moreover, we require the ground state of the theory, namely,  the state  where all the curvatures $R^A$ vanish, to be a particular solution of the equations of motion.
In this configuration, that physically corresponds to the \emph{dynamical vacuum} of the theory, all the 1-form fields are left-invariant  1-forms of the structure group $G$. This last requirement is
 useful in constructing matter coupled or higher dimensional Lagrangians where many fermionic interaction terms are present.
 \item Finally,   all the terms in the supergravity Lagrangian should have the same parity as the Einstein-Cartan term, if we want a parity preserving theory.
 \end{enumerate}
$N=1$, $D=4$ supergravity in absence of matter coupling is particularly simple, and for this theory one easily sees that the only possible term that we can add to the Einstein-Cartan term, fulfilling the requirement of being geometric, together with the other above requirements,
is the Rarita-Schwinger kinetic term. Written in terms of differential forms, it reads:
\begin{align}
\bar\psi\gamma^5\gamma_a\mathcal{D} \psi\,V^a= -i \bar\psi_\mu\gamma^{\mu\nu\rho}\mathcal{D}_\nu\psi_\rho\,\sqrt{g}\,d^4x\,.
\end{align}
Therefore
the Action of $N=1,D=4$ supergravity  must have the following form:
\begin{equation}\label{lagrsuper}
 \mathcal A_{D=4}^{{\mathcal{N}}=1}=  \frac{1}{4\kappa^2}\int_{\mathcal M_4 \subset \mathcal{M}^{[4
 |4]}}\left[R^{ab}\,V^c\,V^d \epsilon_{abcd}+\alpha \overline \psi \gamma_5\gamma_a \mathcal D\psi\, V^a\right]
\end{equation}
 where the coefficient $\alpha$ between the Einstein and the Rarita Schwinger terms is related to the normalization of the gravitino 1-form $\psi$ and will be fixed in a moment.

 Note that the hypersurface  on which the Lagrangian is integrated, $\mathcal{M}_4\subset \mathcal{M}^{[4
 |4]}$,  where the $N=1,D=4$ \emph{superspace} $\mathcal{M}^{[4
 |4]}$ is the base manifold of the principal fiber bundle $[\mathcal{M}^{[4
 |4]},\mathrm{SO}(1,3)]$ with fiber $\mathrm{SO}(1,3)$,
 can be naturally identified with physical space-time. However, as emphasized at the beginning of the present Section, any possible bosonic surface $\mathcal{M}_4$ can be equivalently chosen. Indeed, taking advantage of the fact that our Lagrangian is \emph{geometric}, we know that  the variational principle gives equations independent of the choice of the four dimensional hypersurface $\mathcal{M}_4$.
Note that the fields (1-forms) $ (\omega^{ab},V^a,\psi)$ will depend on all the four bosonic and four  fermionic (Grassmann) coordinates $(x^\mu,\theta^\alpha)$ parametrizing the \emph{superspace}.

 The equations of motion {obtained by varying} $\omega^{ab}{,} \, V^a{}$ and $\psi$, and valid on the full superspace, are, respectively:
\begin{eqnarray}
 \label{eqs1}  \frac{\delta\mathcal{A}}{\delta \omega^{ab}}=0 \ :&& \quad\epsilon_{abcd}\mathcal{D}V^c\wedge V^d + \frac \alpha 4 \overline \psi \gamma_5\gamma_c \gamma_{ab} \psi\, V^c=0\,, \mbox{ that is:}\nonumber\\
 &&\quad \epsilon_{abcd}\left(\mathcal D V^a +\frac{{\ii} \alpha}{8}\,\bar \psi\,\gamma^a\,\psi\right) \wedge V^d =0 {,} \\
\label{eqs2} \frac{\delta\mathcal{A}}{\delta V^a}=0 \ :&& \quad 2 R^{ab}\wedge V^c\epsilon_{abcd}-\alpha \overline \psi \wedge\gamma^5\gamma_d\,\rho =0 {,} \\
 \label{eqs3}
 \frac{\delta\mathcal{A}}{\delta \overline\psi}=0 \ :&& \quad 2\gamma^5 \gamma_a\, \rho \wedge V^a -\gamma_5  \gamma_a\psi \wedge {{T}}^a = 0\,,
\end{eqnarray}
where we used the definition \eqref{curv} to express the gravitino supercurvature  as $\mathcal{D}\psi=\rho$.

  As the equations of motion have to vanish identically when all the (super{-})curvatures are zero (requirement 4.), we see that we must set in the left hand side of equation (\ref{eqs1}) $\alpha =4$ in order to have the super-torsion 2-form $ T^a$ as defined in (\ref{curv}). With this value of $\alpha$, equation  (\ref{eqs1}) takes the form
\begin{equation}\label{tortor}
    T^c \wedge V^d \epsilon_{abcd} = 0
\end{equation}
 and we see that when all the supercurvatures are zero, the equations of motion vanish identically.

To analyze the content of equations (\ref{eqs2}),
(\ref{eqs3}) and (\ref{tortor}), which are 3-form equations  valued on any bosonic hyperplane $\mathcal{M}_4$ immersed  in
of superspace, we expand the curvatures 2-forms  along the basis  $E^{\hat a} \wedge E^{\hat b}$, where $E^{\hat a}=(V^a\,,\psi^\alpha) $ ($a=(0,\dots,3)$ and  $\alpha =1,\dots 4$), of 2-forms in the cotangent space
of superspace, as in \eqref{par}, that is:
\begin{align}
 \label{exptor}
     T^a &= {{{T}}^a{}_{(2|0) bc}} {V^b \, V^c} +   T^a{}_{(1|1)c\,\alpha } \psi_\alpha\,V^c + \psi^\alpha T^a{}_{(0
|2)\alpha\beta} \,\psi^\beta {,}\\
\label{exprho}
    \rho^\alpha &=\rho^\alpha_{(2|0) ab}V^a\,V^b+ \rho^\alpha_{(1|1) a } \psi^\alpha\,V^a + \rho^\alpha_{(0|2) {\beta\gamma}}\psi^\beta\,\psi^\gamma ,\\
    \label{decrab}
   R^{ab}&=   R^{ab}{}_{(2|0) cd}V^c\,V^d +  \overline\Theta^{ab}_c\psi\,V^c+ \bar\psi K^{ab}\psi\,.
\end{align}
In eq. \eqref{exptor}, $  T^a{}_{(1|1)c\,\alpha}$ is a spinor vector and $T^a{}_{(0
|2)\alpha\beta}$ a spinor matrix.  In equation \eqref{decrab}, we have kept for the outer components $R^{ab}{}_{(1|1)}$ and $R^{ab}{}_{(0|2)}$, the names $\overline\Theta^{ab}_c$ and $K^{ab}$ respectively, that were attributed to them originally in the literature.

{We warn the reader that, since we are now in superspace, the rigid indices cannot be traded with  coordinate indices using the bosonic vierbein $V^a_\mu$. Indeed, the full set of supervielbein is now given by $E^{\hat a}= \left(V^a,\psi^\alpha\right)$ and we should invert the matrix $(E^{\hat a}_\mu,E^{\hat a}_\alpha) $ to find the space-time components. For this reason, in the following, we will generally denote  with a tilde the components of the supercurvatures along two bosonic vierbein, that is
$$R^A{}_{(2|0)ab}\equiv \tilde R^A{}_{ab}\,,$$ in order to distinguish them from the space-time projection of the full curvatures.
They are commonly named in the literature as \emph{supercovariant field strengths}.\footnote{The name \emph{supercovariant} means that their supersymmetry transformation law does not contain derivatives of the supersymmetry parameter $\epsilon^\alpha$.}
However, a simpler way to find the space-time components is to project the equations on the space-time basis $dx^\mu\wedge dx^\nu$. For example, from equation  (\ref{exprho}), projecting on the space-time basis we obtain
\textit{} \begin{equation}\label{proj}
 \rho^\alpha _{\mu\nu}=\tilde\rho^\alpha{}_{ ab}V^a_\mu\,V^b_\nu+ \rho^\alpha_{(1|1) a } \psi^\alpha_{[\mu}\,V^a_{\nu]} + \rho^\alpha_{(0|2) {\alpha\beta}}\psi^\alpha_\mu\,\psi^\beta _\nu
 \,, \end{equation}
 where the indices $\mu\nu$ are understood to be antisymmetric. We see that the tilded components of $\tilde\rho_{\mu\nu}$ differ from the real space-time components of $\rho_{\mu\nu}$ by terms in the gravitino fields, namely \emph{outer terms}.
 However, as we will see in a moment,  as far as the $  T^a$ and $\rho$ curvatures of this theory are concerned, we can safely convert rigid Lorentz indices into world indices using the matrix $V^a_\mu $, since in the present case ($N=1$ supergravity in $D=4$) they do not have \emph{outer} components $(V {\wedge} \psi)$ and $ (\psi {\wedge} \psi)$. Then, for the components of the aforementioned curvature 2-forms along $V^a \wedge V^b $ we can neglect the tilde symbol.
 Instead, as we now show and further discuss in the sequel, the distinction is relevant for the Riemann curvuture in superspace, because the space-time projection of the Lorentz curvature does not coincide with its supercovariant field-strength.   }

Let us first work out equation  (\ref{tortor}).
In \eqref{exptor}, the component $ T^a{}_{(1|1)c}$ is a spinor, with $\overline T^a{}_{(1|1)c}$ its adjoint spinor, {while the}  $ T^a{}_{(0
|2)}$ component is a, possibly field-dependent, linear combination  of {gamma matrices}.

  Inspecting the 3-form equation  (\ref{tortor}), where all the components along independent elements of the basis of 3-forms in superspace should vanish independently, one  easily concludes that the components of $ T^a_{(1,1)}$ must be zero,\footnote{The same conclusion can be reached by observing that a spinor scaling as $[L^{-\frac 12}]$ does not exist in the theory.} while, due to the Fierz identity \eqref{3psi41}, $ T^a_{(0,2)}$ could in principle be different from zero  and take the value
  $T^a_{(0,2)}= \beta \gamma^a$, with $\beta$ a free parameter. However, given the definition of the supertorsion, in \eqref{par}, such contribution would only  change the normalization of the gravitino, {so that putting $ T^a_{(0,2)}=0$ just amounts to fixing such normalization.

  In summary, we get that the supertorsion $T^a$ has the following parametrization on a basis of 2-forms in superspace:
  \begin{align}
      T^a= \tilde T^a{}_{ bc} V^b\wedge V^c=T^a{}_{ bc} V^b\wedge V^c\,,
  \end{align}
 precisely as the torsion $\mathring{T}^a$ of Einstein-Cartan gravity, discussed in Section \eqref{EC} (see in particular the  discussion after equation  \eqref{tordef}). It follows that equation  (\ref{tortor}) has exactly the same form, and therefore the same solutions, as in pure gravity case,  provided we replace the  bosonic torsion $\mathring{T}^a$  with the supertorsion $ T^a$ defined in \eqref{curv}. In this way,
with the same computations as those made for pure gravity in Section \ref{EC}, one easily {obtains} the vanishing of the  {$\tilde T^a{}_{bc}=0$}  components and therefore that the {whole super-torsion 2-form} is zero:
 $$T^a=0\,.$$
 For the bosonic case, this equation  is solved by expressing the components of the spin connection as   functions of the vielbein and its space-time derivatives. Note, however, that in the supergravity case, solving for the spin connection $\omega^{ab}_\mu$  with the usual procedure gives a spin connection that depends not only on the bosonic vielbein and their derivatives but also on gravitino bilinears (see e.g. reference  \cite{Castellani:1991et}).

We can now apply the same procedure  to solve the equations (\ref{eqs2}) and (\ref{eqs3}), by expanding   the curvatures  $\rho$ and $R^{ab} $ along a complete basis of 2-forms in superspace, according with \eqref{par}.
  As $ T^a=0$, the equation  (\ref{eqs3})  takes the form
\begin{equation}\label{psivrho}
 2\gamma^5 \gamma_a\, \rho \wedge V^a=0\,,
\end{equation}
which is the superspace expression of the Rarita-Schwinger equation.
 We should now use in equation  (\ref{psivrho}) the expansion of  $\rho\equiv \mathcal D\,\psi$ on a basis of 2-forms in superspace:
\begin{equation}\label{exprho1}
    \rho^\alpha =\rho^\alpha_{(2|0) ab}V^a\,V^b+ \rho^\alpha_{(1|1) a } \psi^\alpha\,V^a + \rho^\alpha_{(0|2) {\beta\gamma}}\psi^\beta\,\psi^\gamma ,
\end{equation}
where, for the sake of clarity, we have made the spinor index explicit. Then, from equation  (\ref{psivrho})  one easily realizes that $\rho^\alpha_{(1|1) a }=\rho^\alpha_{(0|2) {\alpha\beta}}=0$,  so that the 2-form $\rho$ has only components $\rho_{(2,0) ab}\equiv \tilde \rho_{ab}= \rho_{ab}$}, on the cotangent space of $\mathcal M_4$, namely
\begin{equation}\label{gravinocur}
   \rho={ \rho_{ab}} V^a\,V^b= \rho_{\mu\nu}\,dx^\mu\wedge dx^\nu\,.
\end{equation}
If we now project eq. \eqref{psivrho} on space-time, it gives the space-time gravitino  equation  of motion:
\begin{align}   \gamma^5 \gamma_a\, \rho \wedge V^a= 0 \quad \Rightarrow \quad
 \gamma^5 \gamma_a\, \rho_{bc}  V^a_\sigma V^b_\mu V^c_\nu \sqrt{g} d^3 x \epsilon^{\mu\nu\sigma\lambda}=0\end{align}
 that is
 \begin{align}
\epsilon^{\mu\nu\sigma\lambda}\gamma_5\gamma_\sigma\rho_{\mu\nu}=0 \,,\ \mbox{or, equivalently:} \quad\epsilon^{\mu\nu\sigma\lambda}\gamma_5\gamma_\sigma\mathcal{D}_\mu \psi_\nu=0\,.\label{grafi}
\end{align}
This is the Rarita-Schwinger equation, in its standard formulation.

Finally, by expanding  $R^{ab}$ as in \eqref{decrab},
from equation  (\ref{eqs2}) we find \footnote{ To solve for $\Theta_{ab|c }$ one performs the cyclic permutations of the indices $(a,b,c)$ and uses the same trick used in the bosonic case to compute the affine connection in terms of the metric and its derivatives.}
\begin{equation}\label{tetta}
{\overline\Theta^{ab}_c =- \epsilon^{abrs} \bar{ \rho}_{rs}\gamma_5\gamma_c - \delta^{[a}_c\epsilon^{b]mst}\bar{ \rho}_{st}\gamma_5 \gamma_m}
\end{equation}
while $K^{ab}=0$, as it can be also  easily checked by observing that no gamma-matrix-valued object (field or parameter) scaling, in natural units, as $[L^{-1}]$, exists in the pure theory.

   In conclusion{,} the solution of the equations of {motion} (\ref{eqs1}), (\ref{eqs2}), (\ref{eqs3}) for the \emph{outer} and \emph{inner} projections of the curvature multiplet gives:
\begin{eqnarray}
  \label{parcur}  R^{ab}&=& \tilde R^{ab}_{cd}V^c\,V^d + \overline\Theta^{ab}_c\psi\,V^c {,} \\
  \label{partor}  {{T}^a} &=& 0 {,} \\
   \label{parrho} \rho &=& {\rho_{ab}}  V^a\,V^b\,,
\end{eqnarray}
where $\overline\Theta^{ab}_c$ is given by \eqref{tetta}, in which the non-vanishing \emph{outer component} of the Lorentz curvature, $R^{ab}_{(1|1) a}$, is written in terms of the \emph{inner component} of the gravitino curvature, $ \tilde\rho_{ab}$.

Let us  emphasize here some peculiarities of the result found above for the specific model of minimal supergravity  in $D=4$, but which express  general features of supergravity in the geometric approach:
\begin{itemize}
    \item The above parametrizations, equations \eqref{parcur},\eqref{partor},\eqref{parrho}, of the supercurvatures defined in \eqref{curv},  have been obtained as solutions of the field equations, that is they  \emph{hold on-shell}. As we will see later in Section \ref{rheo}, this is related to the fact, already discussed in Section \ref{on}, that \emph{supersymmetry is an on-shell symmetry}.

    As a consequence, the following general rule will hold true in superspace:

    \emph{Off-shell},
    the supercurvatures $R^A(\mu)$  are given in terms of  their definitions \eqref{curv}, while  \emph{after applying the variational principle to the Action}, that is \emph{on-shell}, the supercurvatures will have to satisfy their parametrizations \eqref{par}, that is in particular, for the case under study, equations  \eqref{parcur},\eqref{partor},\eqref{parrho}.
    \item
    As exhibited in \eqref{tetta},  the non-vanishing \emph{outer components} of the supercurvatures $R^A$
  are linearly expressed, on-shell, in terms of
  \emph{inner} components of the set of  supercurvatures $R^A$.
  This  general property is called \emph{rheonomy}.

   Physically, this property guarantees that  the theory in superspace does not include extra on-shell {degrees} of freedom, besides those already present in space-time.

\end{itemize}

 Finally, inserting  the parametrizations  (\ref{parcur}), (\ref{partor}), (\ref{parrho}) in the equations of motion (\ref{eqs1}), (\ref{eqs2}), (\ref{eqs3}),  we get the components  of the equations of motion along {$V^a \, V^b\,V^c$, that is:}
\begin{eqnarray}
   \tilde R^{ac}{}_{bc} -\frac12 \delta^a_b  \tilde R^{cd}{}_{cd}&=& 0 {,} \\
  \tilde T^a_{bc} &=& 0 {,} \\
 \label{spacetimeeq} \epsilon^{abcd}\gamma^5\gamma_c  {\tilde \rho_{ab}} &=& 0 \,.
\end{eqnarray}
Expressing the supercovariant field-strengths in terms of the physical curvatures projected on space-time
\footnote{
We obtain them by projecting the equations  (\ref{parcur}), (\ref{partor}), (\ref{parrho}) on the space-time 2-form differentials $dx^\mu \wedge dx^\nu$.}:
\begin{align}
\label{supercov}  R^{ab}_{\mu\nu}&= \tilde R^{ab}{}_{cd}V^c_\mu\,V^d_\nu + \overline\Theta^{ab}_c\psi_{[\mu}\,V^c_{\nu]}= \tilde R^{ab}{}_{\mu\nu}+ \overline\Theta^{ab}_{[\nu}\psi_{\mu]} {,} \\
   T^a_{\mu\nu} &=  \tilde T^a{}_{bc} V^b_\mu\,V^c_\nu= \tilde T^a{}_{\mu\nu} {,} \label{Tmunu}\\
  \rho_{\mu\nu} &=\tilde\rho_{ab}V^a_\mu\,V^b_\nu =\tilde \rho_{\mu\nu}\,,
\end{align}
we get the space-time field equations.  In particular, we see that the Einstein equation  of motion
contains extra terms linear in the inner components {$\rho_{ab}{\equiv} \rho_{\mu\nu}\,V^\mu_a\,V^\nu_b$}. These {terms} give rise to the energy-momentum tensor of the gravitino field $\psi_\mu$.  Furthermore, we remark that eq. \eqref{Tmunu} implies, in terms of the torsion 2-form $\mathring{T}^a$: $\mathring{T}^a= \frac i2 \bar \psi \gamma^a \psi$.

\subsubsection{Supersymmetry invariance of the Action.}

Let us now check the supersymmetry invariance of the Action \eqref{lagrsuper}, that we rewrite:
\begin{equation}\label{act1}
 \mathcal A_{D=4}^{{\mathcal{N}}=1}=  \frac{1}{4\kappa^2}\int_{\mathcal M_4 \subset \mathcal{M}^{[4
 |4]}}\left[R^{ab}\,V^c\,V^d \epsilon_{abcd}+4 \overline \psi \gamma_5\gamma_a \mathcal D\psi\, V^a\right]\,.
\end{equation}
In the geometric approach, it is expressed by the vanishing of the Lie derivative of the Lagrangian 4-form for infinitesimal diffeomprphisms in the fermionic directions of superspace.}
Using the Lie derivative with a tangent vector
 $  \epsilon= \epsilon^\alpha   D_\alpha${,} where $  D_\alpha$ is the tangent vector dual to $\psi^\beta$, introduced in \eqref{Lieder}, supersymmetry invariance requires:
\begin{equation}\label{bob}
  \delta_\epsilon \mathcal L\equiv \ell_\epsilon \mathcal L={\iota_{ \epsilon}}d\mathcal L + d(\iota_{ \epsilon}\mathcal{L})= 0\,,
\end{equation}
up to boundary terms.
If we impose appropriate boundary conditions on $\mathcal{M}_4$, assuming that the fields  vanish at radial infinity so that any exact form does not contribute to the action, then we may discard the total derivative term {$d (\iota_{\epsilon}\mathcal L)$} and other possible exact 4-forms on the right-hand side. \footnote{The analysis can be extended also to theories with non-trivial boundary conditions, see \cite{Andrianopoli:2014aqa},
\cite{Concha:2018ywv},\cite{Andrianopoli:2020zbl},\cite{Andrianopoli:2021rdk}, even if we will not discuss here such cases.  }
Note that here
$d\mathcal{L}$ is not automatically zero, since the 4-form $\mathcal{L}$ is not a top form in  the {(4+4)-}dimensional superspace.
Taking into account the supercurvature definitions (\ref{curv}) and their Bianchi identities \eqref{bi1}, a simple computation gives
\begin{align}\label{invar}
 d\mathcal L=\mathcal{D}\mathcal{L}= \frac{1}{4\kappa^2}&\Bigl[2  R^{ab}\left( T^c
 +\frac {\ii}2{\bar \psi}\,\gamma^c\psi
 \right)\,V^d\epsilon_{abcd}   +4 \bar \rho\,\gamma^5 {\gamma_a} \rho V^a +\nonumber\\
  &+ \bar \psi\gamma^5 {\gamma_c\,\gamma_{ab}} \psi\,R^{ab}\,V^c  -4 \bar\psi\gamma^5 {\gamma_a} \rho \,\left( T^a
 +\frac {\ii}2{\bar \psi}\,\gamma^a\psi
 \right)\Bigr]\,.
\end{align}
Using the Fierz identity \eqref{3psi41} (see also reference \cite{Castellani:1991et}) and performing some gamma matrix manipulations{,} one is left with:
\begin{equation}\label{dL}
  d\mathcal L=R^{ab}\,  T^c \,V^d\epsilon_{abcd}+ \bar \rho\gamma^5 {\gamma_a} \rho V^a -4\bar\psi\gamma^5 {\gamma_a} \rho {  T^a} \,.
\end{equation}
  Finally{,} contracting with the tangent vector $\epsilon=\epsilon^\alpha  D_\alpha$ along an odd direction of superspace{,} we obtain
\begin{eqnarray}\label{invar2}
 &&\iota_{ \epsilon}\,(d\mathcal L)= 2 (\iota_\epsilon R^{ab})\,  T^c \,V^d\epsilon_{abcd} + 2R^{ab}(\iota_\epsilon\,   T^c) \,V^d \epsilon_{abcd} {+}
   8(\iota_\epsilon \bar \rho)\gamma^5 {\gamma_a}\rho V^a \nonumber\\
  &&- 4\bar \epsilon\gamma^5 {\gamma_a} \rho\,  T^a -4 \bar \psi\gamma_5 {\gamma_a} (\iota_\epsilon\rho)   T^a {-}4 \bar\psi\gamma^5 {\gamma_a} \rho  (\iota_\epsilon \,  T^a).
\end{eqnarray}

From {(\ref{act1})} we see that we can have an invariant action if \begin{align}
 \iota_{ \epsilon}\,(d\mathcal L)   = d( \mbox{3-form})\,,
\end{align}
that is, if we require constraints on the components of the curvatures.

This is obtained if we set
\begin{equation}\label{consrta1}
 \iota_\epsilon T^a=0;\quad \iota_\epsilon \rho =0
\end{equation}
and furthermore
\begin{equation}\label{consrta2}
 2\left(\iota_\epsilon R^{ab}\right)\,{V^d}\,\epsilon_{abcd}-4\bar \epsilon\gamma_5 {\gamma_c} \rho  =0\,,
\end{equation}
in which case we find $\delta_\epsilon \mathcal L =0$, that is, \emph{invariance of the Lagrangian under supersymmetry}.

We note that the requirements  {(\ref{consrta1})}  are trivially satisfied if we use,  for the contraction of the supercurvatures,   the  constraints (\ref{partor}) and (\ref{parrho}),    while {(\ref{consrta2})} is satisfied using the parametrization \eqref{parcur}, which gives:
\begin{align}
    \iota_\epsilon R^{ab}= \overline \Theta^{ab}_c\epsilon V^c\,,
\end{align}
where $ \Theta^{ab}_c$ has been defined in equation  (\ref{tetta}).

We recall that the above constraints on the supercurvatures were found from the equations of motion. In other words, requiring supersymmetry invariance of the superspace Action we retrieve exactly the same constraints on the curvatures as those found from the equations of motion.

 We conclude that the \emph{supergravity Lagrangian is invariant under (local) supersymmetry transformations, when the superspace curvatures are expressed by their on-shell parametrizations}, equations  (\ref{parcur})-(\ref{parrho}).
This in turn implies that the supersymmetry transformations  leaving the Lagrangian invariant do not form a closed algebra, unless one uses the equations of motion.
We remark that, to obtain the result, it was crucial the use of the Bianchi identities \eqref{bi1}, expressing the closure of the supercurvatures in superspace.

Let us now explicitly work out the supersymmetry transformation laws leaving the Action invariant, and check that they close the supersymmetry algebra only on-shell.
We can evaluate them by applying the Lie derivative formula \eqref{Lieder}, to write down the superspace diffeomorphisms of the gauge fields ${\omega^{ab},V^a,\psi}$. We should use a generic tangent vector on the full fiber-bundle. This means including, besides the tangent vectors $D_{{a}}$ and $D_\alpha$ on $\mathcal{M}_{4|4}$,
 dual to the   1-forms $V^a$ and $\psi^{\alpha}$, also
the tangent vector $D_{ab}$ dual to the spin-connection $\omega^{ab}$, that is such that $D_{ab}(\omega^{cd})= 2\delta_{ab}^{cd}$, so that the general form of the parameter is $\vec \epsilon= \frac 12\epsilon^{ab}D_{ab} + \epsilon^a D_{a} +\epsilon^\alpha D_\alpha$. We find:
\begin{eqnarray}
 \delta_\epsilon \omega^{ab} &=&(\nabla \epsilon)^{ab} + \epsilon^c V^d R^{ab}_{cd}+   \overline\Theta^{ab}_c \psi \epsilon^c +   \overline\Theta^{ab}_c \epsilon V^c {,} \\
  \delta_\epsilon V^a &=& (\nabla \epsilon)^{a} {,} \\
  \delta_\epsilon \psi^\alpha  &=& (\nabla \epsilon)^\alpha+ \epsilon^a \rho_{ab}^\alpha V^b.
\end{eqnarray}
Restricting ourselves to the Lie derivative along the fermionic supersymmetry parameter $\epsilon^\alpha$ only,  that is setting $\epsilon^{ab}=\epsilon^{a}=0$, we have
\begin{eqnarray}
\label{susyosp1} \delta_\epsilon \omega^{ab} &=&(\nabla \epsilon)^{ab} + \overline\Theta^{ab}_c \epsilon V^c {,} \\
\label{susyosp2} \delta_\epsilon V^a &=& (\nabla \epsilon)^{a} {,}  \\
 \label{susyosp3} \delta_\epsilon\psi^\alpha  &=&(\nabla \epsilon)^\alpha.
\end{eqnarray}
Here {the} symbol $\nabla$ denotes the \emph{$\overline{\rm O{S}p(1|4)}$ covariant derivative} of the coadjoint multiplet $\mu^A =(\omega^{ab},V^a,\psi)$ of $\overline{\rm O{S}p(1|4)}$, to be distinguished from the \emph{Lorentz} covariant derivative, $\mathcal D$.
Explicitly we find, for the supersymmetry transformations laws of the fields on space-time:
\begin{eqnarray}
\label{finsusyR}\delta_\epsilon \omega^{ab}_\mu &{=}&   \overline\Theta^{ab}_c \epsilon  V^c_\mu\\
\label{finsusyT} \delta_\epsilon V^a_\mu &=&  -{\ii}\bar\psi_\mu \gamma^a \epsilon {,} \\
\label{finsusyro}\delta_\epsilon\psi_\mu  &=& {\mathcal D_\mu \epsilon\,} \,.
\end{eqnarray}

Now we recall that the Lie derivative along tangent vectors $\tilde T _A$ satisfy an algebra isomorphic to the Lie algebra of the vector fields
$[\tilde {T}_A,\tilde {T}_B]=\left(C^A_{\;BC}+R^A_{\;BC}\right)\,\tilde T_C$, namely
\begin{equation}\label{liederalgebra}
  [{\ell}_{\tilde T_A},{\ell}_{\tilde T_B}]=\ell{_{[\tilde T_A, \tilde T_B]}} ,
\end{equation}
if the supercurvatures $R^A_{\;BC}$ are completely general, that is, if they do not satisfy any cons\-traint. In our case they satisfy the constraints (\ref{parcur})-(\ref{parrho}) and, in general, the Lie derivative algebra, namely, the algebra of supersymmetry transformations, \emph{cannot close off-shell}.
This can be checked explicitly by considering the commutator of two supersymmetry transformations on the fields, with parameters $\epsilon_1^\alpha$, $\epsilon_2^\alpha$.  In particular, the above operation on the vielbein gives:
\begin{align}   [\delta_{\epsilon_1},\delta_{\epsilon_2}]V^a_\mu=-i\,\mathcal{D}_\mu \left(\bar{\epsilon_1}\gamma^a \epsilon_2
    \right)\,,
\end{align}
that is it reproduces the local supersymmetry algebra \eqref{osp14},
while on the gravitino the calculation, after some $\gamma$-matrix manipulation gives
\begin{align}   [\delta_{\epsilon_1},\delta_{\epsilon_2}]\psi_\mu = -i \, \left(\bar{\epsilon_1}\gamma^\nu \epsilon_2
    \right)\mathcal{D}_{[\mu} \psi_{\nu]} + \alpha(x)_\mu{}^\lambda \gamma_{\lambda\nu\rho}\rho^{\nu\rho}\,,
\label{delta2psi}\end{align}
where
\begin{align}
    \alpha(x)_\mu{}^\lambda=\frac 18 \left(2\bar{\epsilon_1}\gamma^\nu \epsilon_2A^\lambda{}_{\mu\nu}  + \bar{\epsilon_1}\gamma^{\nu\sigma} \epsilon_2 B^{\lambda}{}_{\mu\nu\sigma}
    \right)\,,
\end{align}
$A^\lambda{}_{\mu\nu},B^{\lambda}{}_{\mu\nu\sigma}$ being some linear combinations of $\gamma$-matrices, whose precise definition, with the details of the calculation, can be found in \cite{Castellani:1991et}, Vol. 2, page 636.
Note, in particular,  that equation  \eqref{delta2psi}  reproduces the supersymmetry algebra only after imposing the gravitino field equation  \eqref{grafi}.

Actually,
requiring that the  Bianchi identities on the constrained curvatures be satisfied, one finds that their components on the bosonic cotangent plane $R^A_{rs}$ satisfy the equations of motion of the theory. It follows that the supersymmetry algebra of the transformations leaving the Lagrangian invariant, associated to the tangent vectors $\epsilon^\alpha\,D_\alpha$,  will in general only \emph{close on-shell}, that is, only if the equations of motion are satisfied.

As a final comment, it is interesting to compare the supersymmetry transformation laws \eqref{finsusyR}, \eqref{finsusyT}, \eqref{finsusyro} with the
$\overline{\rm O{S}p(1|4)}$-gauge covariant derivative of the fields in the adjoint multiplet, with parameter $\vec \kappa= \frac 12\kappa^{ab}J_{ab} + \kappa^a P_{a} +\kappa^\alpha Q_\alpha$ (for a more extended discussion on this point,  see Appendix \ref{Lie}). They read:
\begin{eqnarray}
 \delta^{(gauge)} \omega^{ab} &=&  (\nabla \kappa{)}^{ab}=\mathcal D {\kappa^{ab}} {,} \\
 \delta^{(gauge)} V^a &=& (\nabla \kappa)^{a}=\mathcal D \kappa^{ab} + \kappa^{ab} V_b -{\ii}\bar \psi \gamma^a \kappa {,} \\
  \delta^{(gauge)} \psi&=&(\nabla \kappa)= \mathcal D \kappa -\frac14 \kappa^{ab}\gamma_{ab}\psi {,}
\end{eqnarray}
and, setting again  $\kappa^{ab}=\kappa^{a}=0$ they reduce to the \emph{gauge-supersymmetry transformations}, that is, gauge transformations along odd directions of the structure group. Projected on space-time, they are:
\begin{eqnarray}
\delta_\kappa^{(gauge)} \omega^{ab}_\mu &=& 0 {,} \label{gsusyomega}\\
 \delta_\kappa^{(gauge)} V^a_\mu &=&  -{\ii}\bar \psi_\mu \gamma^a \kappa {,} \\
  \delta_\kappa^{(gauge)} \psi_\mu&=& \mathcal D_\mu \kappa \,.
\end{eqnarray}
We note in particular the difference in the supersymmmetry trasformation of the spin connection, equation \eqref{finsusyR}, with respect to its  gauge-supersymmetry transformation, equation \eqref{gsusyomega}.

Let us now summarize the result obtained so far:
Even if the supercurvatures {${T}^a$} and $\rho$, whose parametrizations are given in  equations (\ref{partor}) and (\ref{parrho}),  respectively,  have no components along the fermionic vielbein $\psi$, a non{-}vanishing component along $\psi {\wedge} V^a$ does appear in the on-shell value of the Lorentz supercurvature, that is, (\ref{tetta}). \emph{This is sufficient to exclude factorization of the odd fermionic coordinates}. Indeed its presence makes the supersymmetry transformation a \emph{diffeomorphism} in superspace and \emph{not a gauge transformation}.

It must also be noted  that  the absence of such fermionic components in the (on-shell) gravitino curvature $\rho$ implies that the supersymmetry variation of $\psi$, given in equation  (\ref{finsusyro}){,}
is the same as if the {Lagrangian} were  invariant under supersymmetry \emph{gauge transformations}.  However, the supersymmetry transformations of the Lagrangian actually correspond to superdiffeomorphisms, which close the supersymmetry algebra only on-shell, and not to gauge transformations
\footnote{Note that if the {Lagrangian} were invariant under supersymmetry gauge transformations the superfields would only depend on the $x^\mu$ coordinates.}. The point is that  such behavior of the gravitino transformation law is due to the very simple form of the minimal ${\mathcal{N}}=1$, $D=4$ pure supergravity. Any other supergravity with ${\mathcal{N}}>1$ or $D>4$ or even the same theory ${\mathcal{N}}=1$, $D=4$ coupled to matter multiplets exhibits a gravitino curvature with components ${\rho_{(1,1)}}\neq 0$ so that the $\delta_\epsilon \psi$ will have, besides the Lorentz covariant derivative of the supersymmetry parameter{,} also terms along {$\psi \wedge V^a$}.

\vskip 3mm

 As an  example, let us consider ${\mathcal{N}}=2$, $D=4$  pure supergravity. Here the supergroup is $\overline{\rm O{S}p(2|4)}$. The coadjoint gauge supermultiplet is now given by $\mu^A=(\omega^{ab}, V^a,\psi_i, \mathcal A)$, where $\mathcal A$ is {a} $\rm U(1)$ gauge field 1-form  and the index $i=1,2$ enumerates the  gravitinos in the two{-}dimensional representation of $\mathrm{U}(2)$.

 The definitions of the associated supercurvatures are obtained by starting from the Maurer-Cartan equations dual to the algebra of the structure supergroup
 {and} deforming the left-invariant 1-forms into {non} left-invariant ones. Without giving the derivation, for each of them we write, besides {the} definitions of the supercurvatures in the first line,  also (in the second line) their on-shell parametrization,  as found from the analysis of the equations of motion:
\begin{eqnarray}
R^{ab} &{\equiv}& d\omega^a_{\,\,\,b}+ \omega^a_{\,\,\,c}\omega^c_{\,\,\,b} \nonumber \\
& =& \tilde R^{ab}_{cd}V^c V^d + \overline\Theta^{ab}_{i|c}{\psi^i} V^c -\bar\psi_i\left(\tilde F^{ab}+\frac{\ii}2 \tilde F_{cd}\epsilon^{abcd} \gamma_5\right)\psi_j {\epsilon ^{ij}} \,, \\
{ T}^a &{\equiv}& \mathcal D V^a -\frac{{\ii}}{2} \bar\psi_i \gamma {\psi^i}\nonumber \\&=&0 \,, \\
F &{\equiv}& d\mathcal{A} +  {\epsilon ^{ij}} \bar\psi_i \psi_j \nonumber\\
&=&\tilde F_{ab} V^a V^b \,, \\
\rho_i &{\equiv}& \mathcal D\psi_i\nonumber\\
&=& \tilde\rho_{i|ab} V^a V^b +\left( \gamma^a \,\tilde F_{ab} + {\ii} \gamma_5 \gamma^a \frac{\ii}2 \tilde F^{cd}\epsilon_{abcd}\right){\epsilon _{ij}\psi^j} V^b \,.
\end{eqnarray}
 The important thing to note is that the parametrization of the {curvature} 2-forms
  are all given in terms of {their inner components}, namely, $\tilde R^{ab}_{cd}$, {$\tilde\rho_{i|ab}$} and {$\tilde F_{ab}$} (${ T}^a_{bc}$ is zero) \footnote{Note that {$\tilde F_{ab}= F_{ab}$}, since the supercurvature {$F$} has components only along $V^a \, V^b$.}.

  Since the on{-}shell value of the supercurvatures is known, the supersymmetry transformation laws of the coadjoint supermultiplet, now containing also $\mathcal A$, can be obtained at once from the general formula (\ref{Liederiv}). Looking at the Lie derivative formula, we see that the transformation laws of the multiplet of fields can be simply obtained  performing the contraction of the on-shell curvatures with respect to the tangent vector $\bar{\epsilon}\,D$ and adding to the gravitino transformation the Lorentz covariant derivative of the supersymmetry parameter $\epsilon^{\alpha i}$, as it happens in the $\overline{\rm O{S}p(1|4)}$ case.  We find:
\begin{eqnarray}
 \label{finsusyR2} \delta_{\epsilon} \omega^{ab} &=& \overline\Theta^{ab}_{i|c} {\epsilon^i} V^c {,} \\
\label{finsusyT2} \delta_{\epsilon} V^a &=&   - {\ii\bar{\psi}_i \gamma^a \epsilon^i} {,} \\
\label{finsusyro2}\delta_\epsilon\psi_i  &=&\mathcal D \epsilon_i+ {\ii} \,\epsilon_{ij}F^{ab} V^b \gamma^a\epsilon_j+ {\ii} \,\frac12\epsilon_{ij}\epsilon_{abcd}F^{cd} V^b \gamma_5 \gamma^a {\epsilon^j} {,} \\
\label{finsusyF}\delta_\epsilon \mathcal A &=& 2 {\epsilon^{ij}} \bar\psi_i\,\epsilon_j {.}
\end{eqnarray}
From this example we see that, in general, not only the Lorentz curvature $R^{ab}$, but also the other supercurvatures have non-vanishing components along the (outer) $\psi$-directions.

\subsubsection{Bosonic Kinetic Terms in the Geometric Approach}\label{kin}
 Let us end this subsection on the supergravity action, with a more detailed explanation of why we cannot admit the Hodge duality operator in the construction of the Lagrangian, and how to remedy its absence when gauge potentials of internal symmetries, and more generally bosonic p-forms, are present.

 This issue is not of academic interest only, since in matter coupled supergravity theories, and also in pure supergravity for theories with more than 4 supercharges ($N>1$ in $D=4$), the spectrum of the theory includes bosonic fields, whose  kinetic terms are quadratic in their field-strengths.

 The standard way to write a quadratic kinetic term  requires the  use of the duality Hodge operator. Indeed, considering a $p$-form potential gauge field $A^{(p)}$ in a theory in $D$ space-time dimensions, its field-strength $F$ is a $(p+1)$-form, $^*F $ its Hodge-dual, and the kinetic term in the Lagrangian $D$-form is (the precise coefficient  depends on $p$ and on the number of space dimensions of the specific theory  considered):
 \begin{align}
     F\wedge {}^* F
     \propto F_{\mu_1 \dots\mu_{p+1}}F^{\mu_1 \dots\mu_{p+1}}\,\sqrt{g}\,d^D x\,.
 \end{align}
 This expression however is background dependent and therefore
 introduces  a dependence of the Lagrangian on the hypersurface $\mathcal M_D$ and its metric.
As such, and also because the Hodge operator critically depends on the dimensionality of the space where it is applied, it also makes problematic the extension of the fields domain from the bosonic hypersurface $\mathcal{M}_D$ (space-time) to the superspace $\mathcal{M}_{D|N}$, and \emph{a fortiori}, to the full structure supergroup.

Actually, the way out from this impasse is very simple. It is sufficient to write the kinetic terms of the boson fields in \emph{first order formalism}.
For example, let us consider  an abelian 1-form gauge field $A=A_\mu\,dx^\mu= A_a V^a$ in $D=4$ space-time, with field-strength $F=dA=\partial_\mu A_\nu dx^\mu\wedge dx^\nu=F_{ab}V^a\wedge V^b$. The standard expression for its kinetic term in the Action will be
\begin{equation}
     \mathcal{A}=-\int F^{\mu\nu}\,F_{\mu\nu}\sqrt{-g}\, d^4x=\frac12 \int F\wedge {}^*F.
\end{equation}
We may avoid use of the Hodge operator ${}^*$, if we introduce an auxiliary  0-form antisymmetric tensor field  $\hat F_{ab}=-\hat F_{ba}$ and write the new  kinetic term as follows;

\begin{equation}
\frac{-1}{4!} \int \hat {F}_{ab}\,\hat F^{ab} \epsilon_{pqrs}V^{pqrs}+\alpha \int\hat {F}^{ab} F\,V^{cd}\epsilon_{abcd} .\\
\end{equation}
Varying  the Lagrangian with respect to  $\hat F^{ab}$ we find that, choosing $\alpha =\frac12$, we obtain

\begin{equation}
    \hat F_{ab}=F_{ab}
\end{equation}
where $F_{ab}$ are the components of the 2-form $F$ along the vierbeins. Varying next with respect to the gauge field $A$ one finds the usual equation  of motion

\begin{equation}
\mathcal{D}_\mu F^{\mu\nu}=\mathcal{D}_a F^{ab}=0\,.
\end{equation}
In this way we see that, adopting a first-order formalism, we can write a \emph{geometric} Lagrangian without using the Hodge duality operator.

Thus, quite generally, \emph{a Lagrangian is geometric if it is constructed in terms of p-forms, wedge products, the exterior derivative $d$ and without the use of the Hodge duality operator}.

\subsection{The principle of Rheonomy}\label{rheo}
We can now resume our analysis of the previos subsection in the following way:

 Supersymmetry can be interpreted geometrically as the  requirement that the superspace Action be invariant under diffeomorphisms along odd directions of superspace, Effectively, this corresponds to the fact that the superspace equations of motion imply that \emph{the outer components of the super-curvatures are expressible algebraically (actually linearly) in terms of the components along two inner vielbein}. As already mentioned, this property has been called \emph{rheonomy}. Note that \emph{rheonomy  is just a geometrical interpretation of supersymmetry originally introduced on space-time}.
  Explicitly, the occurrence of \emph{rhenomy} can be written as follows:
\begin{equation}\label{algebraic}
    R^A_{\alpha\,C}= C^{A|mn}_{\alpha\,C|B}\, R^B_{mn} {,}
\end{equation}
where $C^{A|mn}_{\alpha\,C|B}$ are suitable invariant tensors of  the supergroup ${\mathrm {G}}$ defining the basic superalgebra on which the theory is constructed, ${\mathrm {G}}$=$\overline {\rm O{S}p(1|4)}$  in our case.  The geometric meaning of this property can be better understood if we use the Lie derivative formula (\ref{lie2}) in superspace. Inserting (\ref{algebraic}) in the Lie derivative formula (\ref{lie2}) for a  supergroup ${\mathrm {G}}$ we obtain:
\begin{equation}\label{rhe}
    \delta \mu^A = (\nabla \epsilon)^A + 2{\bar{\epsilon}} \,C^{A|mn}_{\alpha\,C|B}\, R^B_{mn}.
\end{equation}
 On the other hand{,} the Lie derivative can be interpreted either from the \emph{passive} or  from {the} \emph{active} point of view. From the passive point of view, the supersymmetry transformation
 along the $\epsilon^\alpha=\delta \theta^\alpha $ parameter is interpreted as the lift in $\mathcal{M}_{4|4}$, from a given $\mathcal M_4$ to an infinitesimally close $\mathcal M^\prime_4$, which does not change the physical content of the theory, since it is described by the same Lagrangian, after performing a supersymmetry transformation (and  a Lorentz gauge transformation) \footnote{The passive interpretation of the Lie derivative explains the world \emph{rheonomy} given to this geometrical interpretation of supersymmetry. Indeed, referring to the lift $\mathcal M_4 \rightarrow \mathcal M^\prime_4$, in ancient Greek {``}rhein" means flow and {``}nomos" means law{.}}. From the active point of view, however, it transforms a given configuration on $\mathcal M_4$, which we can take as space-time, setting $\theta^\alpha=\delta\theta^\alpha=0$, to another physically equivalent configuration  on the same space-time hypersurface.
 This property allows us  to restrict the theory, the Lagrangian and the equations of motion, to any such arbitrarily chosen  hypersurface $\mathcal M_4 \, (\theta^\alpha=d\theta^\alpha=0)$, embedded in superspace and identified with space-time.

One can  now appreciate why we have illustrated in detail the mechanism of the Lorentz coordinate factorization in the gravity case defined on the Poincar\'e manifold.
Actually the interpretation of the rheonomy mechanism  is quite analogous to the interpretation of Lorentz transformations for gravity constructed directly on a group manifold. Indeed, in the case of pure gravity,  we have seen  that a transfer of information from any $\mathcal M_4$ to any other $\mathcal M^\prime_4$ implies a $\rm SO(1,3)$ transformation or{,} equivalently, a change of Lorentz configuration on the fixed space-time hypersurface.
 On the other hand, in our example of ${\mathcal{N}}=1$, $D=4$ supergravity,
besides  deducing the factorization of the Lorentz coordinates  exactly as  in the pure gravity case, we have further illustrated that the equations of motion allow us to deduce that the  transfer of information concerns not only Lorentz gauge transformations but, what is our main goal, also \emph{supersymmetry}.
Coming back to supersymmetry in the geometric approach, the supersymmetry transformations relate the fields on a given bosonic hypersurface $\mathcal{M}_4\subset \mathcal{M}_{4|4}$, to the fields on any other bosonic submanifold $\mathcal M^\prime_4\subset \mathcal{M}_{4|4}$.
However, the difference between $\rm SO(1,3)$ transformations and supersymmetry is that, due to the horizontality of the curvatures in the  Lorentz directions, the supergroup $\rm G$ acquires the structure of the fiber bundle $[\mathcal{M}_{4|4},\rm SO(1,3)]$. The Lie derivative along Lorentz
 directions in $\tilde G$ amounts to a Lorentz gauge transformation. On the other hand, in the case of supersymmetry,  curvatures are not horizontal along the $\psi$  gauge fields, and the Lie derivative, in this case, gives to supersymmetry the geometric interpretation of superdiffeomorphism.

Finally, we remark that the property of working on any   hypersurface $\mathcal{M}_4$ immersed in superspace and  identified with space-time, without the need of specifying a metric on it \footnote{Clearly different surfaces are just different sections of the principal fiber bundle, and are therefore \emph{locally} equivalent. We will not discuss here global properties of the bundle.}, makes this approach quite different from the ordinary  \commas Supergravity approach"
where the fields of the Lagrangian are expressed in a given set of coordinates $(x^\mu,\theta^\alpha)$. In that approach, they are expanded in the Grassmann-odd coordinates and the integration in superspace is made using the Berezin integration on the Grassman-odd sector.

\subsubsection{{The Role of the Bianchi Identities.}}

Until now we have described how a supergravity Action in superspace can be constructed in the geometric approach and how to find the supersymmetry transformations that leave it invariant in superspace.
We have also shown that in this framework supersymmetry transformations can be given  a geometrical interpretation as (super)diffeomorphisms in superspace.

A crucial role in this construction is played by the structure group $G$, which in the simple case of minimal pure supergravity in $D=4$ is the super-Poincar\'e group $G=\overline{\mathrm{OSp}(1|4)}$, and by the adjoint multiplet of fields $\mu^A$ ($A=1,\cdots, adj(G)$), with its  $G$-supercurvatures \begin{align}
R^A\equiv d\mu^A + \frac 12 C^A{}_{BC}\mu^B\wedge \mu^C\,,\label{curG}\end{align} where $C^A{}_{BC}$ are the structure constants of $G$, that in the dynamical vacuum satisfy the Maurer-Cartan equations of $G$.

They are, therefore, defined by symmetry principles (see the discussion in appendix \ref{mc}), as an invariant set of supercurvatures of $G$ and, as such, the consistency of the theory is encoded in the cohomology of the exterior derivative operator $d$ (in the condition $d^2=0$, which is equivalent to the Jacobi Identities on the structure constants), that is, in their \emph{Bianchi identites} which are obtained by direct application of $d$ to \eqref{curG}:
\begin{align}
 dR^A+ C^A{}_{BC}\mu^B\wedge R^C =0\,. \label{bigen}
\end{align}
Since supersymmetry, as discussed in Section \ref{on}, is an \emph{on-shell symmetry} then, when the curvatures $R^A$ are dynamical supercurvatures of a supergroup,   \eqref{bigen} is not anymore an identity, but becomes a relation that holds only upon use of the field equations. In other words, we can say that the curvatures $R^A$ are  \emph{formally defined } on symmetry arguments, but  their Bianchi \commas identities" \eqref{bigen} are in fact \emph{equations} to be satisfied \emph{on-shell} by the parametrization of the supercurvatures in superspace, equation \eqref{par}, according with  the principle of Rheonomy discussed above.

 This opens an equivalent and powerful approach to the construction of the supergravity theories in superspace (equations of motion and transformation laws), which does not rely on the existence of an Action, but is based on a systematic use of the Bianchi identities \emph{ assuming rheonomy} from the very beginning.

 The Bianchi identities, then, assume the role of differential constraints among the space-time components of the supercurvature parametrizations. These differential constraints, on the other hand, can be nothing else, in disguise, than the equations of motion \footnote{Together with all the other restrictions on the theory required by supersymmetry, among which in particular, when the theory includes scalar fields, the relations characterizing the geometry of the scalar $\sigma$-models.}, since Bianchi identities become identities on-shell, and cannot conflict with the differential equations obtained from the Lagrangian. Once the field equations are obtained in this way{,} the {Lagrangian}, if desired, can be easily reconstructed.
 In the actual computations one usually couples the two methods, namely the Lagrangian approach and the Bianchi  equations, to arrive in the simplest way to the final determination of the parametrization of the curvatures in superspace (and thus to the supersymmetry transformation laws) together with the determination of all terms in the {Lagrangian}.

\section{Higher p-Forms Supergravities and their Hidden Supergroups.}\label{4}

We have often stressed that the mechanism of rheonomy actually holds in all supergravities, independently of the number of supersymmetries, the dimensionality of space-time{,} and their matter couplings, if any. However, apart from few exceptions, most of the higher dimensional theories have a gravitational multiplet containing antisymmetric tensors of rank higher than 1. Similarly,  matter supermultiplets also can have higher rank tensors. In these cases, the group manifold interpretation presented in the previous Sections as a possible starting point for supergravities, whose fields are defined on a group manifold, has to be reconsidered. Indeed the coadjoint multiplet of a (super{-})group consists of 1-forms dual to the group generators, with no room for higher $p$-forms.

In the present Section we will show, referring mainly to the case of {$D=11$} supergravity, where this development was first presented \cite{D'Auria:1982nx}, that
the Maurer-Cartan equations can be generalized to more general structures,  admitting in the set of Maurer-Cartan 1-forms also higher $p$-forms ($p>1$). The resulting generalized Maurer-Cartan equations, satisfying the integrability requirement $d^2=0$, can be seen as a natural extension of {Lie algebras} in their dual formulation and can accommodate  supermultiplets containing higher $p$-forms.

These results were obtained in \cite{D'Auria:1982nx} {by R. D'Auria and P. Fr\'e} for maximal $D=11$ supergravity. The space-time Lagrangian of $D=11$ supergravity theory was previously derived using the Noether approach in reference \cite{Cremmer:1978km}, and includes a 3-index antisymmetric tensor, namely  a 3-form gauge potential, in the gravitational multiplet.

The occurrence of a 3-form in the supergravity multiplet can be easily understood as a consistency condition for the theory to be supersymmetric, which requires the matching of the bosonic and fermionic   on-shell propagating degrees of freedom of the theory.

In eleven-dimensional space-time, the vielbein has on-shell
$\frac 12 D(D-3)=44 $   d.o.f., while the gravitino field has $2^{[D/2-1]}(D-3)= 128$ on-shell d.o.f.. Thus we need 84 more bosonic d.o.f. in order for the bosonic and fermionic d.o.f. to match. They are in fact provided by an on-shell propagating 3-form  potential. Indeed, for a propagating antisymmetric tensor gauge potential of rank three, $A_{\mu\nu\rho}$, we have
$\frac{1}{3!} (D-2)(D-3)(D-4)=84$ d.o.f., so that the requirement is satisfied.

Even if this extra 3-form cannot be interpreted as a dual of a generator of a Lie algebra, nevertheless  the authors of \cite{D'Auria:1982nx} tried  to give a fully geometrical interpretation of the space-time formulation of the theory in such a way that all the nice properties of the geometrical group manifold approach could be extended also to that case.

 In their geometrical approach, the authors of \cite{D'Auria:1982nx} introduced for the first time a generalization of the Maurer-Cartan equations in terms of an integrable systems containing higher $p$-forms, which they called \emph{Cartan Integrable Systems} (CIS).
 In the following years, they  recognized, in a paper of Sullivan \cite{Sullivan} on Free Differential Algebras,  part of the properties of the Cartan Integrable Systems and, for this reason, they
  changed the original name CIS into \emph{Free \-Differential Algebra (FDA)}, which is the name still currently used in the supergravity literature \footnote{Actually, as pointed out in reference \cite {nlab} the FDA denomination is a slight misnomer from the mathematical point of view, since these graded algebras are not \emph{free} as differential algebras, but only \emph{semi-free} (see also footnote \ref{fda}). }.

  When, after several years, these structures went under the scrutiny of mathematicians, see e.g. \cite{nlab}, it was pointed out that the CIS/FDA were, historically, \emph{the first example} of the so-called \emph{$L_\infty $ algebras}, which were introduced in the mathematical literature more than a decade later, albeit in a dual language (See \cite{Lada:1992wc} and references therein; for a comprehensive reference, see \cite{nlab}).\\ By dual language here we mean
 the equivalence
 of two different structures: on one side we have graded Lie algebras
 $\mathfrak{g}$ with an operator of \commas derivation", $D$, acting as a bracket on a couple of vectors and mapping them on a linear combinations of  vectors:
\begin{align}
  D(T_A,T_B)\equiv   [T_A,T_B]=C^C_{\;\;AB}T_C\,;
\end{align}
  on the other side we have instead the dual graded co-algebra over its
 dual vector space $\mathfrak{g}^*$  where the  dual of the \commas derivation"  acts on the 1-forms as the exterior derivative operator $d:d^2=0$:
 \begin{align}
 d\sigma^A+\frac12 C^A{}_{BC}\sigma^B\wedge\sigma^C=0.
 \end{align}
 The equivalence is guaranteed by the equivariance relation:
 \begin{align}
 d\sigma^A(T_B,T_C)= -\frac 12 \sigma^A\left([T_B,T_C]\right)\,.
 \end{align}
Further details can be found in Appendix \ref{mc}.

 As we will discuss in the present Section, the new structure,  introduced in \cite{D'Auria:1982nx}, is  a generali\-zation of the Maurer-Cartan equations to a set  of  higher $p$-forms $\Theta^{A(p)}$,  where the differential nilpotent operator \commas $d$" acts on a given $p$-form by mapping it to a polynomial of the same set.
From a mathematical point of view, such construction, acting on a collection of $p$-forms with $1\leq  p\leq n-1$, was understood as a
 \emph{dualization} of an  $L_n$ algebra. In its standard  formulation, the \commas derivation" operator $D$ dual to the exterior differential
takes the form of a \commas higher bracket" structure of the $L_n$ Lie algebra.
See references \cite{Lada:1992wc},\cite{Sati:2015yda},\cite{nlab} for details on definitions of $L_n$ algebra.
Moreover, the identity $d^2=0$ becomes in the $L_n$ case what is called \emph{the strong homotopy identity}, which must be satisfied in order to have a consistent $L_n $ algebra.
Since this can be done for any $n$, we may say that \emph{the new structure introduced in reference} \cite{D'Auria:1982nx}
 \emph{is a dual formulation of an
  $L_\infty$ algebra} \footnote{ Note that from this point of view the Maurer-Cartan equations are the dual of a $L_2$ algebra and viceversa.}.

It must be said that models reproducing the structure of the $L_{\infty}$ algebras
 also appeared in the physical literature at the beginning of the nineties, more or less when the mathematical definition of $L_\infty$ algebra appeared in the literature (see references \cite{Zwiebach:1992ie,Henneaux:1992ig,deWit:2002vt,Andrianopoli:2007ep,Sati:2008eg}, and \cite{stronghom} for an extended collection of papers on results in Physics with $L_\infty$-algebra type structure.).

In the following, we will give a short account of how to construct such  \commas FDA"'s and how to apply the method to $D=11$ supergravity. Moreover, we will show that, starting from the super Poincar\'e group in eleven dimensions from which the FDA algebra can be obtained,
we can further reduce the FDA to a \emph{hidden} \emph{ordinary Lie graded algebra}.

\subsection{ Cartan integrable Systems as a Generalization of Maurer Cartan equations.}

The essential point of the D'Auria-Fr\'e construction is the possibility of building higher integrable systems, introducing forms of higher degree and mimicking the structure of the Cartan-Maurer equations of a graded Lie algebra.

Indeed, suppose we introduce on a manifold $\mathcal{M}_D$, whose dimension $D$ is not determined for the moment, a set of  $p$-forms $\{\Theta^{A(p)}\}$ of various degrees $1\leq p\leq p_{\mathrm{max}}$,
where $A(p)$ is an index in a given representation of a structure group $G$,
such that their exterior derivatives can be expressed as a polynomial in the set of $\{\Theta^{A(p)}\}$ itself, with constant coefficients:
\begin{equation}
d{\Theta^{A(p)}}+\sum_{n=1}^{N}\,\frac{1}{n!} C^{A(p)}{}_{B_{1}(p_1)B_{2}(p_2)\dots B_n(p_n)}
\Theta^{B_1(p_1)}\wedge\Theta^{B_2(p_2)}\wedge\dots \wedge\Theta^{{B_n}(p_n)}=0\,. \label {deteta}
 \end{equation}
where $N=p_{\mathrm{max}+1}$.
The constant coefficients $C^{A(p)}{}_{B_{1}(p_1)B_{2}(p_2)\dots B_n(p_n)}$ are actually invariant tensors {of $G$}.
Note that, {being $\Theta^{A(p)}$ a $p$-form, then} $p_1+p_2+\dots +p_n= p+1$.

It is also important to stress that the symmetry in the exchange of two lower neighbouring indices of the constant $C$-coefficients is inherited by the exchange of two neighbouring $\Theta^{B(p)}$, namely \footnote{This change of sign in  permuting two contiguous  indices is called Koszul sign law in mathematics.}:

\begin{equation}
B_i(p_i)\,B_{i+1}(p_{i+1}) = (-1)^{|B_i| |B_{i+1}|+ p_i p_{i+1}}B_{i+1}(p_{i+1})B_i(p_i)\,,\label{exchange}
\end{equation}
where $|A(p)|$ denotes the grading of the form $\Theta^{A(p)}$.
Let us now impose the integrability of equation  (\ref{deteta}) namely $d^2=0$:

\begin{align}
 &d^2  {\Theta^{A(p)}}=-\sum_{n=1}^N \frac{1}{(n-1)!} \sum_{m=1}^N \frac{1}{m!}
C^{A(p)}{}_{B_{1(p_1)}B_{2(p_2)}\dots B_{n(p_n)}}\,C^{B_1(p_1)}_{D_{1(q_1)} D_{2(q_2)}\dots D_{m(q_m)}} \times\nonumber \\
& \Theta^{D_1(q_1)}\wedge \Theta^{D_2(q_2)}\wedge\dots \wedge\Theta^{D_m(q_m)}\wedge
\Theta^{B_2(p_2)}\wedge\dots \Theta^{B_n(p_n)}=0 \label{closure}.
\end{align}
The above closure condition is satisfied if the set of invariant tensors $C^{A(p)}{}_{B_{1}(p_1)B_{2}(p_2)\dots B_n(p_n)}$ satisfies the \commas generalized Jacobi identities":
\begin{align}
C^{A(p)}{}_{B_{1(p_1)}\bigl[B_{2(p_2)}\dots B_{n(p_n)}}\,C^{B_1(p_1)}{}_{D_{1(q_1)} D_{2(q_2)}\dots D_{m(q_m)}\bigr]}=0  \,,   \label{kos}
\end{align}
where we denoted by $[...]$ the graded symmetrization of the indices, according to the Koszul sign-law  \eqref{exchange}.

Even if not evident at first sight,  eq. \eqref{closure}, first introduced in '81 in \cite{D'Auria:1982nx}, reproduces, in a dual form, the condition of the \emph{strong-homotopy Jacobi identity} satisfied by an $L_\infty$ algebra, which is usually  formulated in terms of \commas higher brackets". This is shown in detail, for example, in reference
\cite{nlab}, where equation  \eqref{closure}  is rewritten in a way
to make it explicit.
There, one can find several different but equivalent definitions of an $L_{\infty}$ algebra. In particular, the definition of the $L_{\infty}$ algebra, given there in terms of a {\bf{semifree differential graded algebra}},  $\mathfrak{g}^*$, matches the definition, given in the original paper \cite{D'Auria:1982nx}, of CIS/FDA.
The equivalence is simply obtained by passing from the graded vector space $\mathfrak{g}$ of a finite dimension $n$, to its degree-wise dual vector space $\mathfrak{g}^*$ which we may identify with the Grassmann graded vector space of the p-forms $\Theta^{B(p)}$.
In  Appendix \ref{Linfinity}  we will report  some details on
the equivalent formulation of $L_\infty$-algebras in the standard form and how our formalism can reproduce their strong homotopy identity.

Note that in the case that the coalgebra $\mathfrak{g^*}$ is an ordinary Lie algebra in dual form, namely $L_2$ in dual form, this is  referred to as  Chevalley-Eilenberg algebra.
Therefore,  \emph{the  FDA can be considered  a generalization of the Chevalley-Eilenberg algebra from ordinary graded Lie algebras to higher graded Lie algebras of $p$-forms}. In particular, at least when the vector space $\mathfrak{g }$ is finite dimensional and we have an operator   $D$ with $D^2=0$ acting as a derivation, one can pass to the dual graded vector space $\mathfrak{g }^*$ whose Grassmann algebra is naturally equipped with the usual exterior derivative $d$. This gives a semifree differential graded algebra, which reproduces our approach in terms of the so-called FDA.
 As for the equivalence between our FDA and the extension of the Chevalley-Eilenberg to higher $p$-forms, see  \cite{CEalgebra}.

\subsection{The construction of the FDA.}

We show in this Section how to generate a FDA
starting from an ordinary Lie algebra.
We start from the Maurer-Cartan equation  of the Lie (co)algebra $\mathbb{G}$ of a given Lie group $G$, with subgroup $H\subset G$:
\begin{equation}
d\sigma^A +\frac12 C^A_{\;\;BC}\sigma^B\wedge \sigma^C=0.\label{MC}
\end{equation}
Next, we consider an \emph{$H$-orthogonal Chevalley cochains complex} \footnote{By $H$-orthogonal cochain we mean that if the Lie algebra has a symmetry subgroup which is a gauge symmetry of the theory, then the associated $p$-form cannot enter in the construction of the cochain. More  precisely, a  cochain $\Omega^i_{(n,p)}$ is $H$-orthogonal if $\iota_H \Omega^i_{(n,p)}=0=\iota_H \nabla^{n}\Omega^i_{(n,p)}=0$. For example, if the theory we are constructing includes  Lorentz transformations $\rm SO(1,10)$
 (which is a subgroup of the (super)-Poincar\'e group), the gauge field $\omega^{ab}$ cannot enter in the construction of the cochain.
    In a sense, this extends to higher forms the notion of \emph{$H$-factorization} introduced in Section \ref{Sugra} (see equation \eqref{facto}). Note, however, that working with the \emph{relative} Chevalley-Eilenberg algebra we should consider, as derivation operator,  the $H$-\emph{covariant derivative} $\mathcal{D}_{(H)}$, such that $(\mathcal{D}_{(H)})^2= R^{(H)A}T_A$, $R^{(H)A}$ being the $H$-curvature, which  vanishes in the FDA  but is not zero out of the vacuum. \label{relCE} },
 namely polynomials of $p$-forms of the following type:
\begin{equation}
\Omega^i_{(n,p)}=C^{i}_{{A_1},\dots,A_n}\,\sigma^{A_1}\wedge,\dots,\wedge \sigma^{A_n}
\end{equation}
where $i=1,\cdots ,n$ runs in a $n$-dimensional  representation $D^{(n)}(T_A)^i{}_j$ of the Lie algebra generators $T_A$ of  $\mathbb{G}$,   ${A_1,\dots,A_p}$ being indices in the coadjoint representation and  $C^i_{A_1,\dots, A_p}$ constants invariant tensors of $G$.

Next we introduce the $\mathbb{G}$-covariant derivative $\nabla^{(n)}$ acting on the $\Omega^i_{(n,p)}$:
\begin{equation}
(\nabla^{(n)})^i_j={d}\delta^i_j +\sigma^A\wedge D^{(n)}(T_A)^i_j\,.
\end{equation}\label{boundary}
Actually, in the case of  a Maurer-Cartan set of 1-forms, $\nabla^{(n)}$ coincides with the $G$-covariant derivative computed at $R^A=0$ (see appendix \ref{mc}).

Because of equation
\eqref{MC}, we have:
\begin{equation}
\nabla^{(n)}\,\nabla^{(n)}=0
\end{equation}
and as such $\nabla^{(n)}$ is named a \emph{boundary operator}.

If the cochain is closed under $\nabla^{(n)}$, it is a \emph{cocycle}, while a cochain is a \emph{coboundary} if there exists a cochain $\tilde{\Omega}^i_{(n,p-1)}$ such that
\begin{equation}
\Omega^i_{(n,p)} = \nabla^{(n)}\tilde {\Omega}^i_{(n,p-1)}.
\end{equation}
A cocycle which is not a coboundary is a representative of a higher \emph{Chevalley-Eilenberg cohomology class} of the Lie algebra.

Now, given a cocycle $\Omega^i_{(n,p)}$, we can introduce a new form $A^i_{(n,p-1)}$ and write the generalized Maurer-Cartan equation :
\begin{equation}
\nabla^{(n)}\,A^i_{(n,p-1)}+ \Omega^i_{(n,p)}=0,\label{extend}
\end{equation}
also called a \emph{trivialization} of the cocycle.

Adding this equation  to \eqref{MC},  we obtain a \emph{higher Lie algebra}, actually a FDA (\emph{semifree graded differential algebra}).

Of course, the process can be iterated by considering a new set of cochains
 containing, besides the $\sigma^A$, also the $A^i_{(n,p-1)}$, namely:

\begin{align}
&\hat \Omega^{i}_{(n,p)}[\sigma,A]= C^i_{A_1,\dots,A_r\, i_1,\dots,i_s}
\times \sigma^{A_1},\dots,\sigma^{A_r}\wedge A^{i_1}_{(n_1,p_1)},\dots,A^{i_s}_{(n_s,p_s)}.
\end{align}
If we can find new cocycles, say $\Omega^{\prime}$, in this enlarged cochain system, we then have  an enlarged FDA.

The process terminates when no new cocycles can be found so that we have constructed the most general FDA  derived from the Lie algebra.

In the next Section we apply this process to the construction of the general FDA of the eleven dimensional  supergravity, by starting from its underlying Lie algebra, namely the Lie algebra of the super-Poincar\'e group $\overline{{\rm Osp}(32|1))}$. \footnote{With the overline we mean the Inon\"u-Wigner contraction of the $\mathrm{OSp}(32|1)$ to the super-Poincar\'e group. }

\subsection{The FDA associated to the super-Poincar\'e
  Algebra
in {D=11.}}\label{6.1}

The Maurer-Cartan equations of the D=11 super-Poincar\'e graded Lie Algebra are given, in their dual form, in terms of
the set of 1-forms $\sigma^A=(\omega^{ab}, V^a,\Psi^\alpha)$ (with $a,b,\dots=0,1,\cdots 10$, $\alpha=1,\cdots ,32$), where $\omega^{ab}$ is the $\mathrm{SO}(1,10)$ spin connection  and $E^{\hat a}=(V^a,\Psi^\alpha)$ the supervielbein of $D=11$ superspace $\mathcal{M}_{11|32}$, $\Psi$  being a spinor in the 32-dimensional representation  of ${\rm Spin}(32)$. They read:
\begin{align}
 & d\,\omega^{ab}-\omega^a_c\wedge \omega^{cb}=0, \label{poinc1}
\\
 &\mathcal{D}V^a -\frac{i}{2}\bar\Psi\,\Gamma^a\wedge \Psi=0\,.\label{poinc2}
\\
 &\mathcal{D}\Psi\equiv d\Psi-\frac14 \Gamma_{ab} \,\omega^{ab}\wedge \Psi=0 \label{poinc3}\,.
\end{align}
In \eqref{poinc2}, $\mathcal{D}V^a= dV^a-\omega^{ab} \wedge V_b$  and $\mathcal{D}\Psi$ denote the Lorentz covariant derivative of the bosonic and fermionic vielbein respectively.
Because the cohomology is $H$-orthogonal with respect to $H= \mathrm{SO}(1,10)$, the Chevalley cochains can be constructed using only  the supervielbein $V^a,\Psi$.

Let us consider the trivial representation $D^{(0)}$, such that ${\nabla^{(0)}}$
reduces to the exterior derivative $d$. Constructing the Chevalley cohomology one finds that there is a non-trival cocycle of order four, namely:

 \begin{equation}
\Omega_{(V,\Psi)}=\frac12 \bar\Psi\wedge \Gamma^{ab}\Psi\wedge V^a\wedge V^b.
\end{equation}
Indeed
\begin{align}
& d\Omega = \frac{i}{2}\bar\Psi\wedge\Gamma^{ab}\Psi\,\bar\Psi \wedge\Gamma_a\Psi\wedge V^b =0
\end{align}
where we have used equations \eqref{poinc2}, \eqref{poinc3} and the Fierz identity:
\begin{equation}\label{fierz11a}
\bar\Psi\wedge \Gamma_{ab} \Psi\wedge \bar\Psi \Gamma^a\Psi=0,
\end{equation}
which was proven in \cite{D'Auria:1982nx} \footnote{The Fierz identity \eqref{fierz11a} expresses the fact that, in the symmetric product of four Spin(32) representations, the $\mathrm{SO}(1,10)$-vector representation is absent.}.

According to the procedure previously explained, we can therefore  introduce a 3-form $A^{(3)}$ which locally \commas trivializes" the cocycle, writing:
 \begin{equation}
    dA^{(3)}-\frac12 \bar\Psi \wedge \Gamma^{ab} \Psi \wedge V^a \wedge V^b=0\,,\label{da3}
  \end{equation}
  where the factor $1/2$ is our choice of the normalization of $A^{(3)}$.
This equation, added to the Maurer-Cartan equations \eqref{poinc1},\eqref{poinc2} and \eqref{poinc3},
gives a FDA suitable for a geometrical construction of the eleven-dimensional Supergravity.
Indeed, recalling what was said in the preamble of the
 present Section, we see that the just introduced 3-form $A^{(3)}$ gives exactly the d.o.f. necessary to match bosonic and fermionic degrees of freedom of the $D=11$ supergravity theory.

Now, after including $A^{(3)}$ in the enlarged set of MC forms, we can iterate the procedure in order to look for other non trivial cocycles.
We find that there is another cohomology class of order seven given by:
 \begin{equation}
\Omega^\prime (V,\Psi,A)=\frac{i}{2} \,\bar\Psi \Gamma^{a_1,\dots,a_5}\wedge \Psi \wedge V^{a_1}\wedge\dots\wedge V^{a_5}+\frac{15}{2}\bar\Psi \wedge \Gamma^{ab}\Psi \wedge V^a\wedge A^{(3)}.
\end{equation}
This allows us to introduce a 6-form  $B^{(6)}$ locally \commas trivializing" the new cocycle $\Omega^\prime$:

\begin{align}
& d B^{(6)}=\frac{i}{2} \bar\Psi\, \Gamma^{a_1\dots a_5}\wedge \Psi \wedge V^{a_1}\dots \wedge V^{a_5}
 +\frac{15}{2} \bar\Psi \wedge \Gamma^{ab}\,\Psi \wedge V^a \wedge V^b \wedge A^{(3)}.\label{db6}
\end{align}
It can be verified that no new non trivial cocycles can be found. Therefore equations \eqref{poinc1}-\eqref{poinc3} together with \eqref{da3} and \eqref{db6} define  the most general FDA in superspace associated to the eleven dimensional super-Poincar\'e Lie Algebra, whose generators are the MC 1-forms $\sigma^A=(\omega^{ab},V^a, \Psi)$, which are 1-forms of the Lie algebra of $\overline{\mathrm{OSp}(1|32)}$, together with the 3-form $A^{(3)}$ and the 6-form $B^{(6)}$.

\subsection{Geometrical Construction of D=11 Supergravity}\label{6.2}
\par Physical applications of the FDA require a generalization of the concept of left-invariant 1-forms (the set of MC forms $\sigma^A$ in the above subsection) to  non left-invariant \commas soft forms" $\mu^A$, with their associated higher $k$-form curvatures $R^{A(k)}$. The set of 1-forms $\mu^A$ are dynamical fields living on the principal fiber bundle $[\mathcal{M}_{11|32},\mathrm{SO}(1,10)]$, thus  extending  the  Maurer-Cartan equations out of the dynamical vacuum of a supergravity theory (see Appendix \ref{mc}). To construct the $D=11$ supergravity theory, they have to be supplemented by the dynamical 3-form field $A^{(3)}$ satisfying, in the dynamical vacuum, equation \eqref{da3} and by the 6-form $B^{(6)}$ satisfying, in the vacuum, equation \eqref{db6}. The set of dynamical fields is then given by the forms:
\begin{align}
    \Pi^{A(p)}= \left(\omega^{ab},V^a,\Psi,A^{(3),}
    B^{(6)}\right)\,,
\end{align}
 which are in one-to-one correspondence with the left-invariant forms $\Theta^{A (p)}$. To build up the theory, all the concepts advocated for the geometrical construction of supergravity Actions based on Maurer-Cartan equations, can be straightforwardly extended to theories based on FDA's.
One first introduces the (super)-curvatures $R^{A(p+1)}$ of the  $p$-forms $\Pi^{A (p)}$, corresponding to the deviation from zero of equations \eqref{deteta}, when the set of $ \Theta^{A (p)} $ is replaced by the \commas soft" forms $ \Pi^{A (p)} $. Therefore instead of equation  \eqref{deteta} we have:
\begin{equation}
R^{A(p+1)}\equiv d\Pi^{A(p)}+\sum_{i=1}^N \,\frac{1}{n!} C^{A(p)}{}_{B_{1(p_1)}B_{2(p_2)}\dots B_{n(p_n)}}
\Pi^{B_1(p_1)}\wedge\Pi^{B_2(p_2)}\wedge\dots \wedge\Pi^{B_n(p_n)} \label {detetax}
 \end{equation}
and applying the exterior derivative to this equation  we find the generalized Bianchi identity:
\begin{align}
& \nabla R^{A(p+1)}=dR^{A(p+1)}-\sum_{i=1}^N \,\frac{1}{(n-1)!} C^{A(p)}{}_{B_{1(p_1)} B_{2(p_2)}\dots B_{n(p_n)}}\times\nonumber\\
&\times R^{B_1(p_1+1)}\wedge \Pi^{B_2(p_2)}\wedge\dots \wedge \Pi^{B_n(p_n)}=0. \label{detetax1}
 \end{align}
In complete analogy to what one does for 1-forms, equation \eqref{detetax1} defines  the \emph{coadjoint covariant derivative} of the set of $(p+1)$-form field-strengths.

Let us now write down the complete set of differential equations defining the FDA in D=11:
\begin{align}
  \label{defcur1} R^a_{\;\;b} {\equiv} &d\omega^a_{\,\,\,b}- \omega^a_{\,\,\,c} \wedge\omega^c_{\,\,\,b}\,, \\
 \label{defcur2} T^a {\equiv}& \mathcal{D}V^a -\frac{{\ii}}{2} \overline \Psi \Gamma^a \wedge \Psi \,, \\
 \label{defcur3}    \rho {\equiv}& \mathcal D \Psi {= d \Psi - \frac{1}{4} \omega^{ab}}\wedge \Gamma_{ab} \Psi \,, \\
  \label{f4} F^{(4)}\equiv &dA^{(3)} -\frac12\,\overline\Psi\Gamma^{ab}\wedge \Psi\wedge V^a\wedge V^b \,, \\
 F^{(7)}\equiv & dB^{(6)}-\frac{i}{2} \overline\Psi \Gamma^{a_1\dots a_5} \wedge \Psi \wedge V^{a_1,\dots,\wedge V^{a_5}}
 -\frac{15}{2} \overline\Psi \wedge \Gamma^{ab}\Psi \wedge V^a \wedge V^b \wedge A+ \nonumber\\
 &- 15 F^{(4)}\wedge A^{(3)}\,.
   \label{f7}  \end{align}
   The last term in\eqref{f7} (which is obviously zero in the vacuum) has been added to the right-hand side of equation \eqref{f7} in order to have gauge invariance of the curvatures under the higher-form transformations:
   \begin{align}
   A^{(3)}&\rightarrow A^{(3)}+d\phi^{(2)} \,,  \label{gaugea} \\
   B^{(6)}&\rightarrow B^{(6)}+d\lambda^{(5)}
   \label{gaugeb} \,,
   \end{align}
   where $\phi^{(2)}, \lambda^{(5)}$ are general 2-forms and 5-forms respectively.

As previously done for the  theories based on ordinary Lie algebras, to construct a supergravity Action, given the definitions above, one then requires  that:
\begin{itemize}
\item The Action is given in terms of a 11-form Lagrangian   integrated over an eleven dimensional bosonic submanifold $\mathcal{M}_{11}$, immersed in the full superspace $\mathcal{M}_{11|32}$ parametrized by 11 bosonic and 32 fermionic coordinates, $(x^\mu;\theta^\alpha)$, respectively.
\item
The Lagrangian is completely \emph{geometric} that is it is constructed in terms of $p$-forms and wedge products only, without the use of the Hodge-duality operator. In this case it is easily  seen that the fundamental properties of geometric Lagrangians in superspace  based on ordinary Lie algebras,   discussed in some detail in  Section  \ref{Sugra} for the case of pure $D=4$ supergravity, still hold
for more general theories: In particular, even if  the Lagrangian is integrated on a bosonic eleven dimensional hypersurface of superspace (space-time), its being \emph{geometric} gives equations of motion valid on the full superspace.
\item We also add, for physical reasons, some symmetry conditions: the requirement that the Lagrangian be \emph{gauge invariant} under the gauge symmetries of the theory, which include the Lorentz $\mathrm{SO}(1,10)$ gauge symmetry, together with the higher-form gauge invariances, equations \eqref{gaugea} and  \eqref{gaugeb}. Moreover, we add the obvious requirement that all terms scale and have the same parity properties as the Einstein-Cartan term. \footnote {The scaling of each field is immediately obtained from the Maurer-Cartan equations from which also  the scaling of the cocycles is derived.}
\item
It is also useful, as a consistency check on the superspace geometric Lagrangian, to verify that
 its equations of motion admit, among their solutions,  the vacuum solution, namely all the curvatures $R^{A(p+1)}=0$.
\end{itemize}
We notice that the presence of the 6-form $B^{(6)}$, and of the associated curvature $F^{(7)}$ in the FDA, seems to violate the matching between bosonic and fermionic on-shell propagating d.o.f. However,
once the supersymmetry  and gauge invariant Lagrangian have been written down, one finds that all the terms involving the 6-form  $B^{(6)}$ sum up to a total differential and therefore the field $B^{(6)}$ is not propagating. Furthermore, from the  the analysis of the Bianchi \commas identities" in superspace, it also follows that the components along the bosonic vielbein of
the two field strengths $F^{(7)}_{a_1,\dots, a_7}$ and $F^{(4)}_{a_1,\dots, a_4}$ are actually Hodge dual to each other and therefore dynamically the degrees of freedom of $F^{(7)}_{a_1,\dots, a_7}$ are not independent from the ones of $F^{(4)}_{a_1,\dots, a_4}$. Physically, this means that once projected on space-time through $V^a_\mu$, the 7-form field strength $F^{(7)}_{\mu_1,\dots, \mu_7}$, is the \commas magnetic" Hodge dual of the \commas electric" field strength $F^{(4)}_{\mu_1,\dots, \mu_4}$.
\footnote{The above remark shows that the information on the Bianchi \commas identities" in superspace is in general richer than the one available at the Lagrangian level. This holds in particular
when the theory includes mutually Hodge-dual fields.
In this case, it is  possible to write a dual Lagrangian, where $B^{(6)}$, but not  $A^{(3)}$, is among the dynamical propagating \commas electric" fields, its space-time Hodge-dual $A^{(3)}$ being in that case \commas magnetic".}

The obvious conclusion of this Section would be now to construct the D=11 theory. However, we do not report in this contribution the explicit construction of the $D=11$ Lagrangian and/or the associated rheonomic parametrizations  of the graded curvatures, satisfying on-shell the Bianchi identities in superspace.
The explicit construction of the D=11 Supergravity, using these geometric tools,  can be found in references \cite{D'Auria:1982nx} and \cite{Castellani:1991et} (Vol 2, pag 861).

Our interest, in this contribution, has been instead  to show the basic geometric structures for its actual construction, namely, the  structure of its FDA which, as we have discussed (see Appendix \ref{Linfinity}), has an  equivalent description in terms of $L_\infty$ algebras.

\section{FDA and Hidden Lie Algebra of D=11  Supergravity.}
\label{hidden}

An  interesting development of
the geometric  approach in terms of FDA is the following:

\emph{It is possible  to reduce the FDA
of $D=11$ supergravity,
constructed by starting  from the   super-Poincar\'e Lie algebra, in terms of an ordinary graded Lie algebra of which the Poincar\'e algebra is a contraction}.

This was  shown  in the same paper \cite{D'Auria:1982nx}.
There, the authors asked themselves whether one could trade the FDA structure on which the theory is based with a new ordinary Lie superalgebra, written in its dual Cartan form, that is in terms of 1-form gauge fields valued in non{-}trivial tensor and spinor representations of the Lorentz group $\mathrm{SO}(1,10)$. This would {allow to} disclose the fully extended Lie superalgebra hidden in the supersymmetric FDA.

This was proven to be true, and the hidden superalgebra underlying the FDA of $D=11$ supergravity was presented for the first time.

To arrive at the desired result, it was shown that it is possible to associate, to the 3-forms $A^{(3)}$ and the 6-form $B^{(6)}$, a set of bosonic 1-forms $B_{ab}$ and $B_{a_1 \cdots a_5}$ valued  in the antisymmetric representations of $\rm{SO}(1,10)$, and furthermore  an extra \emph{spinor} 1-form $\eta$, in the same spinor representation as $\Psi$. The Maurer-Cartan equations satisfied by the new 1-forms are:
 \begin{eqnarray}\label{news}
 \mathcal D B_{a_1a_2} & = & \frac{1}{2}\overline{\Psi}\wedge\Gamma_{a_1a_2}\Psi \,, \\
\mathcal D B_{a_1...a_5}& = & \frac{i}{2} \overline{\Psi}\wedge \Gamma_{a_1...a5}\Psi\,,\\
 \mathcal D \eta & = & i E_1 \Gamma_a \Psi \wedge V^a + E_2 \Gamma^{ab}\Psi \wedge B_{ab}+ i E_3 \Gamma^{a_1...a_5}\Psi \wedge B_{a_1...a_5}\,, \label{Deta}
\end{eqnarray}
$\mathcal D$ being the  Lorentz-covariant derivatives, and $E_1,E_2,E_3$ some costant coefficients.

The whole consistency of this approach requires:
\begin{itemize}
\item
The $d^2$ closure of the
newly introduced 1-form fields  $B_{ab}$, $B_{a_1\cdots a_5}$ and $\eta$, which are thus included in the  Maurer-Cartan set:
 \begin{align} (\omega^{ab},V^a,\Psi, B_{ab},B_{a_1\cdots a_5},\eta)\,.\label{hid}
 \end{align}
Given the Lorentz-horizontality, in this case it is convenient to consider the relative $\rm {SO}(1,10)$ Chevalley-Eilenberg cohomology, using as derivation operator,  instead of $d$, the Lorentz-covariant derivative $\mathcal{D}$, which at zero curvatures (and in particular,  for $R^{ab}=0$) satisfies $\mathcal{D}^2=0$ (see footnote \ref{relCE}).
 For the two bosonic 1-form fields $B_{ab}$ and $B_{a_1\cdots a_5}$, the  $\mathcal{D}^2$ closure is obvious in the vacuum state, because of the vanishing of the curvatures $R^{ab}$ and $\rho=\mathcal D \Psi$, see \eqref{hid}, while $\mathcal D^2\eta=0$ requires the further condition:
\begin{equation}\label{integrability11}
E_1+10E_2-720E_3=0\, ,
\end{equation}
which can be derived by differentiation and use of the Fierz identities of the wedge product of three gravitino 1-forms in superspace.

\item
An appropriate \emph{parametrization} of the 3-form $A^{(3)}$ on the set of 1-forms spanning the hidden algebra. The most general decomposition of the 3-form in terms of product of the 1-forms \eqref{hid} is
  \footnote{ We do not include the $\omega^{ab}$ connection since  we are using the relative  Chevalley-Eilenberg cohomology.}:
\begin{align}
A^{(3)}_{par}=& T_0 B_{ab} \wedge V^a \wedge V^b + T_1 B_{a b}\wedge B^{b}{}_{c}\wedge B^{c a}
+ T_2 B_{b_1 a_1...a_4}\wedge B^{b_1}{}_{b_2}\wedge B^{b_2 a_1...a_4}+\nonumber\\  &+T_3\, \epsilon_{a_1...a_5 b_1...b_5 m}B^{a_1...a_5}\wedge B^{b_1...b_5}\wedge V^m + \label{a3par}\\
&+T_4 \epsilon_{m_1...m_6 n_1...n_5}B^{m_1m_2m_3p_1p_2}\wedge B^{m_4m_5m_6p_1p_2}\wedge B^{n_1...n_5} +  \nonumber\\
& + i S_1 \overline{\Psi}\Gamma_a \eta \wedge V^a + S_2 \overline{\Psi}\Gamma^{ab}\eta \wedge B_{ab}+ i S_3 \overline{\Psi}\Gamma^{a_1...a_5}\eta \wedge B_{a_1...a_5}\,,\nonumber
\end{align}
where  $ T_i$ ($i=1,\cdots ,5$) and $ S_j$ ($j=1,2,3$) are numerical coefficients.
\item
To show the equivalence of the FDA with this new ordinary super-Lie algebra (in dual form), it is then required that differentiation of 
equation \eqref{a3par} gives the same result as the differentiation of $\mathcal{A}^{(3)}$,
equation \eqref{da3}, namely:
\begin{align}
    dA^{(3)}_{par}-\frac 12 \overline\Psi\wedge \Gamma_{ab}\Psi \wedge V^a\wedge V^b=0\,.\label{da3par}
\end{align}
\end{itemize}
Performing the differentiation and using
 Poincar\'e algebra, together with the differentials of  $B_{ab}, B_{a_1a_2\dots a_5}$ and $\eta$, one finds that the requirement is satisfied provided the $T_i$ and $ S_i $ coefficients are fixed as given in Appendix \eqref{coeff11}. Note that they can be all written in terms of the ratio $E_3/E_2$.
   \footnote{In \cite{D'Auria:1982nx}, the first coefficient $T_0$ was arbitrarily fixed to $T_0=1$ giving only 2 possible solutions for the set of parameters $\{T_i,S_j,E_k\}$. It was pointed out later in \cite{Bandos:2004xw, Bandos:2004ym} that this restriction can be relaxed thus giving a more general solution in terms of one parameter. Indeed, as observed in the quoted reference, one of the $E_i$ can be reabsorbed in the normalization of $\eta$, so that, owing to the relation (\ref{Deta}), we are left with one free parameter, say $E_3/E_2$ Then, in \cite{Andrianopoli:2017itj}, a physical interpretation to the free parameter was given. }.

We remark that the parametrization \eqref{a3par} provides a \emph{trivialization} of the 3-form $A^{(3)}$ of the FDA in terms of the 1-forms defining, in the dual basis, the hidden superalgebra of the theory.
  We stress that to obtain such consistent solutions, \emph{the extra terms
  involving the  spinor 1-form $\eta$ turn out to be necessary:} The Ansatz (\ref{a3par}), if the set extra 1-forms is restricted to the bosonic 1-forms only, does not work. In other words, the inclusion of the spinor 1-form field $\eta$ enters in the decomposition of the 3-form $A^{(3)}_{par}$ in an essential  way, to properly give to $A^{(3)}_{par}$ a decomposition compatible with equation  \eqref{da3}, which describes the FDA on ordinary superspace.

In this way{,} one arrives at the following  set of Maurer-Cartan equations for the left-invariant 1-forms  $(\omega^{ab}\,, V^a\,,\Psi\,, B_{a b}\,, B_{a_1 \ldots a_5}\,,\eta)${:}
\begin{eqnarray}
 \label{deta1} d\omega^{ab}&=& \omega^{ac}\wedge \omega_c^{\;b} ,\\
  \label{deta2}\mathcal{D} V^a &=& \frac{\ii}{2}\overline{\Psi}\wedge \Gamma^a \Psi, \\
 \label{deta3}\mathcal{D} \Psi&=&0, \\
  \label{deta4}\mathcal{D} B_{a_1a_2} & = & \frac{1}{2}\overline{\Psi}\wedge \Gamma_{a_1a_2}\Psi , \\
 \label{deta5}\mathcal{D} B_{a_1 \ldots a_5}& = & \frac{\ii}{2} \overline{\Psi}\wedge \Gamma_{a_1 \ldots a_5}\Psi,\\
 \label{deta6} \mathcal{D} \eta & = & \ii E_1 \Gamma_a \Psi \wedge V^a + E_2 \Gamma^{ab}\Psi \wedge B_{ab}+ \ii E_3 \Gamma^{a_1 \ldots a_5}\Psi \wedge B_{a_1 \ldots a_5}\,.
\end{eqnarray}
{This set of Maurer-Cartan equations identifies (in dual form) the
hidden super-Lie algebra underlying the FDA of $D=11$ supergravity, when the set of  MC forms is extended to include the 3-form  $A^{(3)}$ (but  disregarding $B^{(6)}$), that is:
\begin{align}
 & \label{defcur1'} R^a_{\;\;b} {\equiv} d\omega^a_{\,\,\,b}- \omega^a_{\,\,\,c} \wedge\omega^c_{\,\,\,b} {,} \\
 & \label{defcur2'} T^a {\equiv} \mathcal{D}V^a -\frac{{\ii}}{2} \overline \Psi \Gamma^a \wedge \Psi {,} \\
 &\label{defcur3'}    \rho {\equiv} \mathcal D \Psi {= d \Psi - \frac{1}{4} \omega^{ab}}\wedge \Gamma_{ab} \Psi \, \\
 & \label{f4'} F^{(4)}=dA^{(3)} -\frac12\,\overline\Psi\Gamma^{ab}\wedge \Psi\wedge V^a\wedge V^b\,.
   \end{align}

Let us finally write down the hidden superalgebra in terms of its generators
\begin{equation}
T_{{A}} \equiv\{P_a, Q, J_{ab},  Z^{ab}, Z^{a_1 \ldots a_5}, Q'\}\,,\label{t11d}
\end{equation}
closing a set of graded commutation relations.

They  are dual to the 1-forms $\left(V^a,\,\Psi,\,\omega^{ab}\,,B_{ab},\, B_{a_1 \ldots a_5},\,\eta\right)$ respectively. In particular:
\begin{align}
&\omega^{ab}(J_{cd})= 2\delta^{ab}_{cd}\,,\quad V^{a}(P_b)= \delta^a_b\,,\quad \Psi^\alpha(Q_\beta)=\delta^\alpha_\beta\,,\end{align}
as in $D=4$ supergravity, and furthermore:
\begin{align}
   & B^{ab}(Z_{cd})= 2\delta^{ab}_{cd}\,,\quad B^{a_1\cdots a_5}(Z_{b_1\cdots b_5})= 5!\delta^{a_1\cdots a_5}_{b_1\cdots b_5}\,,\quad \eta^\alpha(Q'_\beta)=\delta^\alpha_\beta\,.
\end{align}
One then finds, as shown in \cite{D'Auria:1982nx} that the $D=11$ FDA which includes the 3-form $A^{(3)}$ among the set of MC forms, corresponds to the following hidden  superalgebra, which can be referred to,  after the authors, as  \emph{DF-algebra}:
\begin{align}
  [J_{ab},J_{cd}]&= -2\,\eta_{a[c}J_{d]b}+2 \,
    \eta_{b[c}J_{d]a}\,,\nonumber\\
[J_{ab},P_{c}]&=-2\,P_{[a}\eta_{b]c}\,,\nonumber\\
\left \{Q,\overline Q\right\} &= -\left({\ii} \Gamma^a P_a + \frac12 \Gamma^{ab}Z_{ab}+ \frac {{\ii}}{5!} \Gamma^{a_1 \ldots a_5}\,Z_{a_1 \ldots a_5}\right)\,, \nonumber\\
\left[ Q',\overline Q' \right] &= 0\,, \nonumber\\
[Q, P_a] &= -2 {\ii} E_1 \Gamma_a Q'\,,\nonumber\\ [Q, Z^{ab}] &=-4 E_2 \Gamma^{ab}Q' \,, \nonumber\\
[Q, Z^{a_1 \ldots a_5}] &=- 2 \,(5!) {\ii} E_3 \Gamma^{a_1 \ldots a_5}Q'\,, \label{hiddenalg}\\
[J_{ab}, Z^{cd}]&=-8 \delta^{[c}_{[a}Z_{b]}^{\ d]}\,,\nonumber\\
[J_{ab}, Z^{c_1\dots c_5}]&=- 20 \,\delta^{[c_1}_{[a}Z^{c_2\dots c_5]}_{b]}\,, \nonumber\\ [J_{ab}, Q]&=- \Gamma_{ab} Q\,,\nonumber
\\
 [J_{ab}, Q']&=- \Gamma_{ab} Q'\,.\nonumber
\end{align}
All the other graded commutators  vanish.

  In the Lie algebra \eqref{hiddenalg}, the  generators {$Q'$, $Z^{a_1 \ldots a_5}$ and $Z^{ab}$ are \commas quasi-central", in
 the sense that they commute with all the algebra but Lorentz generators. They provide a  (quasi)-central extension of the supersymmetry algebra.

 Actually, as shown in reference \cite{Andrianopoli:2017itj}, from a cohomological point of view, to reproduce the integrability of $dA^{(3)}$  the presence of the 1-form $B_{a_1...a_5}$ in the decomposition (\ref{da3})  is not necessary, since all the terms where it appears sum up to give an exact 3-form.
 \footnote{This feature pairs an analogous result for $B^{(6)}$ in the supergravity Lagrangian in superspace, where all the contributions in $B^{(6)}$ sum up to a topological term, as it was shortly discussed at the end of Section \ref{6.2}. }
 {More precisely, in reference \cite{Andrianopoli:2017itj} it was shown that, once formulated in terms of its hidden superalgebra of 1-forms, $A^{(3)}$ can be actually decomposed into the sum of two parts having different group-theoretical meaning:
 \begin{align}A^{(3)}_{par}& = A^{(3)}_{(0)} + \alpha A^{(3)}_{(e)}\, ,\end{align}
where $\alpha$ is a free parameter, and:
\begin{eqnarray}
dA^{(3)}_{(0)}&=&\frac 12 \overline \psi \wedge \Gamma_{ab} \psi \wedge V^a \wedge V^b\label{da30},\\
dA^{(3)}_{(e)}&=&0\label{da3e}\,.
\end{eqnarray}
The part $A^{(3)}_{(0)}=A^{(3)}_{(0)}(V^a,B_{ab},\Psi,\eta)  $, which gives the non-trivial contribution to the 4-form cohomology in superspace,  does not depend on $B_{a_1\cdots a_5}$, while $A^{(3)}_{(e)}= A^{(3)}_{(e)}(V^a,B_{ab},B_{a_1\cdots a_5},\Psi)$ does not contribute to the 4-form cohomology, being a 3-cocycle  of the FDA; however, it enjoys  invariance under  a symmetry algebra which is a parallelization of the (uncontracted) superalgebra $\mathfrak{osp}(1|32)$. It is actually the only contribution in $A^{(3)}_{par}$
  depending on the 1-form $B^{a_1\cdots a_5}$.

 This provides a physical meaning to the free parameter in the solution to equation \eqref{da3par}. More details on this point can be found in \cite{Andrianopoli:2017itj}.
 }

\subsection{D=11 Supergravity and M-theory.}

Several years after the publication of \cite{D'Auria:1982nx}, on the basis of different considerations the same algebra, but \emph{without the inclusion of the  nilpotent generator $Q'$}, was rediscovered. This superalgebra, actually a contraction of the hidden superalgebra (\ref{hiddenalg}),  was named $M$-algebra \cite{deAzcarraga:1989mza,Sezgin:1996cj,Townsend:1997wg,Hassaine:2003vq,Hassaine:2004pp}.
The crucial commutation relation in the M-algebra is the third of \eqref{hiddenalg}:
\begin{align}
 \left \{Q,\overline Q\right\} &= -\left({\ii} \Gamma^a P_a + \frac12 \Gamma^{ab}Z_{ab}+ \frac {{\ii}}{5!} \Gamma^{a_1 \ldots a_5}\,Z_{a_1 \ldots a_5}\right)\,,\label{susyz}
\end{align}
which expresses the anticommutator of two supersymmetry generators, and includes on its right-hand side, besides the translation generator $P_a$, also the quasi-central generators $Z_{ab}, Z_{a_1 \ldots a_5}$.  It is indeed  the natural extension to $D=11$ supergravity of the centrally extended supersymmetry algebra of \cite{Witten:1978mh} (where the central generators were  associated with electric and magnetic topological charges) and{,} as such, has in fact a topological meaning.  The important role of the quasi central generators   $Z_{ab}, Z_{a_1 \ldots a_5}$ was in fact understood  in several papers from  the mid eighties on, see in particular  \cite{Achucarro:1987nc}, \cite{deAzcarraga:1989mza},\cite{Abraham:1990nz},  where it was clarified that they should be associated with extended objects  (2-brane and 5-brane {charges, respectively}), topological defects  in eleven dimensional  superspace.
After the discovery of D$p$-branes as non-perturbative  sources for the R-R gauge potentials \cite{Polchinski:1995mt}, and the following   \emph{second string revolution}, where the role of   duality relations,  and in particular the one between eleven dimensional supergravity and Type IIA string theory in ten dimensions, was clarified,   equation \eqref{susyz} was revived once more. Indeed, the bosonic generators $Z^{ab}, Z^{a_1 \cdots a_5}$ were interpreted as $p$-brane charges, sources of the dual potentials $A_{(3)}$ and $B_{(6)}$ \cite{Hull:1994ys,Townsend:1995gp}, and analogous extended algebras governing the different perturbative descriptions, in space-time dimensions $D\leq 10$, of non-perturbative superstring theory were given. To this structure was then given the name of \emph{M-theory}, explaining why   equation \eqref{susyz} was then referred to as \emph{M-algebra}.

{The $M$-algebra} is now commonly considered as the super{-}Lie algebra underlying $M$-theory \cite{Schwarz:1995jq,Duff:1996aw,Townsend:1996xj} in its low energy limit corresponding to {$D=11$} supergravity in the presence of non-trivial $M$-brane sources \cite{Achucarro:1987nc,Bergshoeff:1987cm,Duff:1987bx,Bergshoeff:1987qx,Townsend:1995kk,Townsend:1995gp}.
Together with its lower dimensional versions, it is understood as the natural generalization of the supersymmetry algebra in higher dimensions, in the presence of non-trivial topological extended sources (black $p$-branes).

 However, if we hold on the idea that the low energy limit of $M$-theory, and then the M-algebra, should be based on the ordinary superspace spanned by the supervielbein $(V^a,\Psi)$, as in the original formulation of {$D=11$} supergravity \cite{Cremmer:1978km},  then the $M$-algebra cannot be the final answer, since it does not contain the extra 1-form $\eta$ dual to the nilpotent generator  fermionic generator $Q'$. Indeed,
 as shown  in \cite{Andrianopoli:2016osu},
  a field theory based on the {$M$-algebra} (but excluding $\eta$, that is setting to zero $Q'$ in \eqref{hiddenalg}) is naturally described on a domain corresponding to an \emph{enlarged superspace} whose cotangent space is spanned, besides the supervielbein $(V^a,\Psi)$, also by the bosonic fields $\{{B_{ab}, B_{a_1\ldots a_5}} \}$, that is in a theory different from 11-$D$ supergravity.
 In order to reproduce the FDA \eqref{defcur1'},\eqref{defcur2'},\eqref{defcur3'},\eqref{f4'} on which {$D=11$} supergravity is based, the presence of $\eta$ among the 1-form generators is necessary. Actually, the DF-{algebra} (\ref{hiddenalg})
differs from its contraction, the $M$-algebra, precisely because it  also \emph{includes  the nilpotent {} fermionic generator $Q'$, ($Q'^2=0$),} dual to the spinor 1-form $\eta$. Indeed, as we have seen in the previous subsection, such spinor 1-form is crucially introduced in the trivialization of the 3-form, equation \eqref{a3par}, in order for equation \eqref{da3par} to hold. Its contribution to the Maurer-Cartan equations of {the} DF-algebra \eqref{deta1}-\eqref{deta6} is given in equation  {(\ref{deta6})}.
As it was shown in \cite{Andrianopoli:2016osu},  equation \eqref{da3par}  in turn implies that the group manifold generated by the DF-{algebra} gets a fiber bundle structure $[\mathcal{M}_{11|32}, \mathcal{H}]$, whose base space is ordinary superspace $\mathcal{M}_{11|32}$, while the fiber $\mathcal{H}\supset \mathrm{SO}(1,10)$ is generated by the subalgebra $\mathfrak{h}$ of the DF-algebra spanned by $(J_{ab},Z_{ab},Z_{a_1\cdots a_5})$. Its cotangent space
is  then spanned, besides the Lorentz spin connection $\omega^{ab}$ of $\mathrm{SO}(1,10)$, also by the bosonic 1-form generators {$B_{ab}, B_{a_1,\dots a_5}$}. Considering the group manifold generated by the DF{-}algebra,  whose   coadjoint multiplet is {$\mu^A=(\omega^{ab}, B_{ab},B_{a_1 \dots a_5},V^a,\Psi, \eta)$},
allows to think of the   1-forms {$B_{ab}$} and {$B_{a_1\ldots a_5}$} as gauge fields in ordinary superspace instead of  additional vielbeins of an enlarged superspace, that is, their curvatures on the fiber are \emph{horizontal}. This is due to the dynamical cancellation of their unphysical contributions to the supersymmetry and gauge transformations with the supersymmetry and {gauge transformations of $\eta$ \footnote{As observed in \cite{Andrianopoli:2016osu}, all the above procedure of enlarging the field space to recover a well defined description of the physical degrees of freedom is strongly reminiscent of the BRST-procedure, and the behavior of $\eta$ is such that it can be actually thought of as a ghost for the 3-form gauge symmetry, when the 3-form is parametrized in terms of 1-forms.}. The same then should apply to the field equations, where the dynamics of all the unphysical contributions is expected to be decoupled from the physical one.

\vskip 3mm
Let us conclude with a final remark. We wonder if the DF-algebra does reproduce the full hidden symmetry of the low-energy, supergravity limit of M-theory, or if some extra generators (maybe an infinite number) have to be included. To our knowledge, the general answer   is still an open problem. We expect, however, that the DF algebra has to be further extended if one wants  to take into account the full non perturbative description of the theory, including the dual Lagrangian description where the 6-form $B^{(6)}$ is electric, $A^{(3)}$ being instead magnetic \footnote{We recall that the space-time projections of the corresponding field-strengths, equations \eqref{f7},\eqref{f4}, are related by Hodge-duality, as discussed at the end of Section \ref{6.2}.}. To disclose the full algebra, in the same spirit of the way opened in \cite{D'Auria:1982nx}, one should find a trivialization $B^{(6)}_{par}$ also for the 6-form $B^{(6)}$, in terms of 1-form fields, such that the FDA relation \eqref{db6} be satisfied by it:
\begin{align}
    d B^{(6)}_{par}=\frac{i}{2} \overline\Psi\, \Gamma^{a_1\dots a_5}\wedge \Psi \wedge V^{a_1}\dots \wedge V^{a_5}
 +\frac{15}{2} \overline\Psi \wedge \Gamma^{ab}\,\Psi \wedge V^a \wedge V^b \wedge A^{(3)}_{par},\label{db6par}
\end{align}
analogously to the prescription \eqref{da3par} for $A^{(3)}$.

This is  still an unsolved issue, also due to the technical complexity of the calculation in the expansion of a 7-form in superspace.
However,
a partial answer was given in \cite{Andrianopoli:2016osu},
where a dimensional reduction of $D=11$ supergravity on the orbifold $T^4/\mathbb{Z}_2$ to the minimal $D=7$ supergravity was considered. In this case, the theory has a rich FDA structure which includes, besides the supervielbein and spin-connection, also a 3-form $B^{(3)}$, with its Hodge-dual form $B^{(2)}$, together with a triplet of 1-forms $A^x$, with their    Hodge duals $A^{(4)|x}$. The hidden algebra trivializing the  mutually Hodge-dual forms $B^{(3)}$ and $B^{(2)}$ was explored in detail in \cite{Andrianopoli:2016osu}, showing that in this case \emph{two inequivalent nilpotent spinor charges are required} to get the hidden algebra, with a fiber-bundle structure and the superspace $\mathcal{M}_{7|8}$ as base space. However, it was also found that two subalgebras of the hidden algebra exist, each of them including only one nilpotent spinor charge. One of the two subalgebras is the relevant one to fully describe the trivialization of $B^{(3)}$, the other, instead, gives the   parametrization in terms of hidden 1-forms of its Hodge-dual $B^{(2)}$. For this reason they were named \emph{Lagrangian subalgebras}.

This analysis suggests that the full hidden algebra of the FDA underlying $D=11$ supergravity should at least include one more spinor charge, playing a role in the hidden description of $B^{(6)}$.
This is left to future investigations.

\section{Acknowledgements}
 We  acknowledge  useful discussions with Urs Schreiber.
 We also thank our friend  and colleague Mario Trigiante for  interesting suggestions and comments.
 \clearpage
 \begin{center}
     Prof. Veeravalli Seshadri Varadarajan: A memory from one of the authors
 \end{center}

I am particularly grateful to the editors R. Fioresi and M. A. LLed\'o for the  chance of honoring  the figure of the eminent
 Mathemathician  V.S. Varadarajan to whom I was related by scientific admiration and a lasting friendship.

I met him for the first time during my frequent visits in the nineties to UCLA University, as a consultant and as a teacher in some PHD topics in group theory and string theory.
There, I also had the opportunity to follow some of his lectures and I immediately  appreciated his way of presenting some issues relevant to physicists. I could appreciate in particular his work on the mathematical aspects
of theoretical physics, as also testified by his interesting book on supersymmetry. We had, actually, an important collaboration, together with S. Ferrara and M. A. Lled\'o, on Spinors Algebras, a topic very useful in supersymmetric theories.

However, the human side of his personality is not less important than his excellent achievements in Mathematics. Immediately after we met for the first time, a friendship was born between us as an effect of our discussions on classical music, to which we shared the same passion.
 Particularly,  we shared a particular admiration for Mozart music. Actually, he was able to reproduce, being an excellent clarinet executor, some Mozart pieces of music for this instrument. Our common interest was  the beginning of a lasting and deep friendship,  which was strengthened  every time I was in UCLA and any time he visited my  Department at Torino Politecnico.   On such occasions, we had several dinners together discussing, besides scientifical topics, also musical events and  literature. He was indeed a man of excellent culture, and I owe to him, among others, the discovery of excellent  indian writers of english language.

 Even when we where separated by the ocean, we had frequent E-mails regarding our  opinion about  some  events concerning new Mozart discographic executions.

This was the kind of our friendship that, even if initially  born from common interests, revealed through the years his gentle character and deep humanity.
Therefore I was much troubled when some years ago he passed away.

This contribution, in collaboration with Laura Andrianopoli, is dedicated to his memory.

\begin{flushright}
Riccardo D'Auria
\end{flushright}

\clearpage

\appendix
\section{Notations and conventions}\label{notations}
All over the paper, we adopted a \commas mostly minus" signature for the space-time metric, that is:
$$
\eta_{ab}= \mathrm{diag}(+,-,\cdots ,-)\,.
$$

\subsection{Fierz identities  in D=4 minimal theory}

The four-dimensional Dirac matrices are defined as
\begin{equation}
    \gamma^a \equiv \begin{pmatrix} \sigma^a & 0 \\ 0 & \overline\sigma^a \end{pmatrix}\,, \quad \gamma_5 \equiv -\frac{i}{4!}\epsilon_{abcd}\gamma^a\gamma^b\gamma^c\gamma^d\,, \quad \gamma^{a_1\dots a_k} \equiv \gamma^{[a_1}\dots\gamma^{a_k]}
\end{equation}
and fulfil
\begin{equation}
     \gamma_0^\dagger = \gamma_0, \hspace{0.5cm}
     \gamma_0\gamma^a\gamma_0 = (\gamma^a)^\dagger,
\end{equation}\begin{equation}
     \gamma_5^\dagger = \gamma_5, \hspace{0.5cm}
    \gamma_5^* = \gamma_5, \hspace{0.5cm}
     (\gamma_5)^2 = \mathbb{I},
 \end{equation}
 \begin{equation}
    \{\gamma^a,\gamma^b\} = 2\eta^{ab}, \hspace{0.5cm}
     [\gamma^a,\gamma^b] = 2\gamma^{ab}, \hspace{0.5cm}
    \gamma^a\gamma^b = \eta^{ab} + \gamma^{ab}\,.
\end{equation}

The charge-conjugation matrix $C$ has the following properties:
\begin{equation}\begin{split}
    C^2=-1\,&, \quad C^T=-C\,, \quad (C\gamma_a)^T=C\gamma_a\,, \quad (C\gamma_5)^T=-C\gamma_5\,, \\
    &(C\gamma_5\gamma_a)^T= -C\gamma_5\gamma_a\,, \quad (C\gamma_{ab})^T=C\gamma_{ab}
\end{split}\label{Eq:ChargeConjMatrix}\end{equation}

The gravitino 1-form in $D=4$ is a Majorana spinor, satisfying the condition:
\begin{align}
    \bar\psi\equiv \psi^\dagger\gamma^0 =\psi^t C\label{psibar}
\end{align}
where $\bar\psi$ denotes the adjoint of the spinor $\psi$.

The following 3-gravitini Fierz identity holds on $D=4$, $N=1$ superspace:
\begin{align}
    \gamma_a \psi \bar\psi \gamma^a \psi=0\,.\label{3psi41}
\end{align}
\subsection{Fierz identities in D=11}
The content of this appendix is taken from \cite{D'Auria:1982nx} and \cite{Andrianopoli:2016osu}.

The gravitino 1-form $\Psi_\alpha $, $(\alpha =1,\cdots , 32)$,  of eleven dimensional supergravity is a Majorana   spinor belonging to the spinor representation of  ${\rm SO(1,10)}$, $Spin(32)$.

The symmetric product
$(\alpha , \beta , \gamma)\equiv\Psi_{(\alpha} \wedge \Psi_\beta \wedge \Psi_{\gamma )}$, whose dimension is $\mathbf{5984}$, belongs to the three-times symmetric, reducible representation  of $Spin(32)$.
Its decomposition into irreducible representations  of $Spin(32)$ gives the 3-$\Psi$  Fierz identities.
One obtains:
\begin{eqnarray}
\mathbf{5984} \to \mathbf{32}+\mathbf{320}+\mathbf{1408}+\mathbf{4224}
\end{eqnarray}
and the corresponding irreducible spinor representations of the Lorentz group $SO(1,10)$ will be denoted as follows:
\begin{equation}
\Xi^{(32)} \in \mathbf{32} \,,\quad \Xi^{(320)}_a \in \mathbf{320}\,,\quad \Xi^{(1408)}_{a_1a_2}\in \mathbf{1408}\,,\quad \Xi^{(4224)}_{a_1...a_5}\in \mathbf{4224}\,,
\end{equation}
where the indices $a_1\cdots a_n$ are antisymmetrized, and each of them satisfies $\Gamma^a \Xi_{ab_1\cdots b_n}=0$.
One can easily compute the coefficients of the explicit decomposition into the irreducible basis, obtaining: \cite {Castellani:1991et}, \cite{D'Auria:1982nx}:
\begin{eqnarray}
\Psi \wedge \overline{\Psi} \wedge \Gamma_a \Psi & = & \Xi^{(320)} _a+ \frac{1}{11}\Gamma_a \Xi^{(32)},  \\
\Psi \wedge \overline{\Psi} \Gamma_{a_1 a_2}\Psi & = & \Xi^{(1408)}_{a_1a_2}-\frac{2}{9}\Gamma_{[a_2}\Xi^{(320)}_{a_2]}+\frac{1}{11}\Gamma_{a_1 a_2}\Xi^{(32)},  \\
\Psi \wedge \overline{\Psi}\wedge \Gamma_{a_1...a_5}\Psi & = & \Xi^{(4224)}_{a_1...a_5}+2 \Gamma_{[a_1 a_2 a_3}\Xi^{(1408)}_{a_4a_5]}+ \frac{5}{9}\Gamma_{[a_1...a_4}\Xi^{(320)}_{a_5]}-\frac{1}{77}\Gamma_{a_1...a_5}\Xi^{(32)} .
\end{eqnarray}

\section{Maurer-Cartan equations and curvatures}\label{mc}
Let us consider a (possibly graded) Lie group $G$, with tangent space spanned by the Lie algebra $\mathfrak{G}$ \footnote{In the discussion here we will consider explicitly the case of a bosonic Lie algebra, but the generalization to the case of a superalgebra is conceptually straightforward.}.
Let
$\{T_A\}$ be the generators of $\mathfrak{G}$, with $A=1,\cdots ,\text{dim}\,\mathfrak{G}$, and commutation relations \begin{align}
    [T_A,T_B]=C^C{}_{AB} T_C\,,
\label{alg}\end{align}
whose consistency rely on the Jacobi identities
\begin{align}
 [[T_{A},T_B],T_{C}]+[[T_{B},T_C],T_{A}]+[[T_{C},T_A],T_{B}]=0
\end{align}
implying:
\begin{align}
     \label{JI}  C^L{}_{[AB} C^D{}_{C]L}=0 \,.
\end{align}
The same algebra can be expressed, in a dual way, in terms of left-invariant 1-forms $\sigma^A$, spanning a basis of the cotangent space of the group manifold $G$ (also called Maurer-Cartan 1-forms, to be referred to, in the following, also as  MC 1-forms) so that
\begin{align}
    \sigma^A(T_B)=\delta^A_B
\end{align}
and satisfying the Maurer-Cartan equations:
\begin{align}
    d\sigma^C+\frac 12 C^C{}_{AB} \sigma^A \wedge \sigma^B=0\,.\label{MC1}
\end{align}
Here, $d$ denotes the exterior differential operator on $G$, which carries 1-form degree. In the dual form of the algebra, the consistency condition is encoded in the cohomological condition $d^2=0$, indeed:
\begin{align}
    0=&d^2\sigma^C= -\frac 12 C^C{}_{AB} d\left(\sigma^A \wedge \sigma^B\right)\nonumber\\
    =&-C^C{}_{AB} d\sigma^A \wedge \sigma^B =-\frac 12C^C{}_{AB}C^A{}_{LM}\sigma^L\wedge\sigma^M\wedge \sigma^B
\end{align}
whose validity implies  \eqref{JI}.
The equivalent dscription  of the algebra in the Maurer-Cartan  formulation  with the one in the standard form, equation  \eqref{alg}, further requires to define the action of the 2-form $d\sigma^C$ on a couple of tangent vectors, as follows:
\begin{align}
 d\sigma^C\left(T_L,T_M\right)= -\frac 12 \sigma^C  \left([T_L,T_M]\right)= -\frac 12 C^C{}_{LM} \,.
\end{align}
The 1-forms $\sigma^A$ can be thought of as the components of the algebra-valued 1-form (pure gauge):
\begin{align}
    \sigma\equiv g^{-1}dg= \sigma^A T_A \in \mathfrak{G}\,,\quad g= \exp{\alpha^AT_A}\in G\,,
\end{align}
$\alpha_A $ being group parameters.
Indeed, then:
\begin{align}
d\sigma =& dg^{-1}\wedge dg= -g^{-1}dg\wedge g^{-1}dg\nonumber\\
=& -\sigma\wedge\sigma
\end{align}
that is, in components:
\begin{align}
d\sigma^C T_C =&  -\sigma^A\wedge\sigma^B T_A \cdot T_B= -\frac 12\sigma^A\wedge\sigma^B [T_A, T_B]= -\frac 12C^C{}_{AB}\sigma^A\wedge\sigma^B T_C
\end{align}
Locally, close to the origin in $G$  (where $g\approx \mathbb{I}+\alpha^AT_A$), $\sigma$  is approximated by an exact form: $\sigma \approx d\alpha^A T_A$.

In physical applications, it is often useful to generalize the notion of MC 1-forms to non left-invariant 1-forms, $\mu$, behaving as $G$-\emph{connections}  on a given base manifold $\mathcal{M}(x)$, interpreted as space-time, of the fiber-bundle structure $[\mathcal{M},G]$. We will sometimes refer to them in the text as \commas soft forms". They can be defined as:
\begin{align}
    \mu(g,x)= g^{-1}\mathring\mu(x) g + g^{-1}dg\,,
\end{align}
where $\mathring{\mu}$ is a $\mathfrak{G}$-valued 1-form on $\mathcal{M}(x)$, and they do not satisfy the Maurer-Cartan equations \eqref{MC1}, but instead:
\begin{align}
 R(x,g)&\equiv   d\mu +\mu\wedge \mu\label{rcomp}\\
 &= g^{-1}\left[d\mathring\mu(x) +\mathring\mu(x)\wedge \mathring\mu(x)\right] g= g^{-1}\mathring R(x) g=R^A(x,g) T_A\,. \nonumber
\end{align}
The quantity
\begin{align}\mathring R(x)\equiv d\mathring\mu(x) +\mathring\mu(x)\wedge \mathring\mu(x)\,,
\end{align}
expressing the failure of the 1-forms $\mathring \mu=\mathring \mu^A(x)T_A $ to satisfy the MC equations,  is an algebra-valued 2-form on $\mathcal{M}$, the   curvature (or field-strength), and it is a tensor in the co-adjoint representation of $G$.
In components, it reads:
\begin{align}
  \mathring{R}(x)=  \mathring{R}^C(x) T_C =\left(d\mathring\mu^C +\frac 12 C_{AB}{}^C\mathring\mu^A\wedge \mathring\mu^B
  \right) T_C \,.\label{r0comp}
\end{align}

If we now expand the 2-form  \eqref{rcomp} on a basis of 1-forms in $G$, that is:
\begin{align}
    R^A(x,g)=  R^A{}_{BC}(x,g)\mu^B \wedge \mu^C\,,
\end{align}
then the expression \eqref{rcomp} can be rewritten in the suggestive, equivalent form:
\begin{align}
d\mu^C +\frac 12 \left[C^C{}_{AB}-2 R^C{}_{AB}(x,g)\right]\mu^A\wedge \mu^B=0
\,,\label{strucfun}
\end{align}
which shows that the non left-invariant 1-forms $\mu^A$ satisfy a would-be MC equation , but in terms of \emph{structure functions} on space-time:
\begin{align}
    \mathcal{C}^C{}_{AB}(x)\equiv C^C{}_{AB}- 2R^C{}_{AB}(x,g)\,,
\end{align}
instead of the \emph{structure constants}
$C^C{}_{AB}$.

\section{ Gauge Transformations versus Diffeomorphisms.}\label{Lie}

We would like  to show here, in an explicit way, how a diffeomorphism  reduces to a gauge transformation when the curvatures are horizontal, while it differs by curvature terms in the general case. We perform the derivation in a general group{-}theoretical setting so that it may apply to any (softened ) group or supergroup $\rm\tilde G$, locally equivalent to a Lie group $G$, that is to any fiber bundle with the group $G$ as its fiber.

An infinitesimal element  of $\rm\tilde G$ is given by a tangent vector on  $\rm\tilde G$, $\vec t=\epsilon^M T_M$, with $ \epsilon^M=\delta x^M {,} $
where the middle alphabet {L}atin capital indices are coordinate indices on  $\rm\tilde G$.
Using the vielbein $\mu^A$ of the whole (soft) group  $\rm\tilde G$ we can rewrite a tangent vector $\vec t$ as follows:
\begin{equation}\label{rewrite}
 \epsilon = \epsilon^{A} \tilde{T}_A,
\end{equation}
where $\epsilon^A= \epsilon^M \mu^A_M$  , and  $\tilde{T}_A= {T}_M \,\mu^M_A$. Here $\tilde{T}_A$ is the vector field generator dual to the non left-invariant 1-form $\mu^A$, $\mu^A(\tilde T_B)=\delta ^A_B, $ and $\epsilon^A = \delta x^A$ is the infinitesimal parameter associated to the shift.
An  infinitesimal
diffeomorphisms generated by $\epsilon^A$ is given by the Lie derivative
\begin{eqnarray}
{\ell}_{\epsilon} \mu^A &=& \left(\iota_\epsilon d + d \iota_\epsilon \right) \mu^A = \nonumber \\
&=& \iota_\epsilon d \mu^A + d \left( \iota_{\left(\epsilon\right)} \mu^A \right)  \nonumber \\
&=& \iota_\epsilon d \mu^A + d \epsilon^A \ .\label{lie}
\end{eqnarray}
where $\iota_{{\epsilon}}$ is the contraction operator along $\epsilon= \epsilon^B\tilde{T}_B$.\\

Adding and subtracting $C^A_{\phantom{A}BC} \mu^B \wedge \mu^C$ to $d \mu^A$ and using the definition of the covariant derivative
\begin{equation}\label{coventry}
  \nabla \epsilon^A = d\epsilon^A +C^A_{\phantom{A}{BC}}\mu^B\,\epsilon^C,
\end{equation}
 we find :
\begin{equation}\label{Liederiv}
{\ell} _{\epsilon} \mu^A = \iota_\epsilon \left(d \mu^A + \frac{1}{2}C^A_{\phantom{A}BC} \mu^B \wedge \mu^C\right) - \epsilon^B C^A_{\phantom{A}{BC}} \mu^C + d \epsilon^A.
\end{equation}
where we have used the antisymmetry of $C^A_{\phantom{A}{BC}}$ in the lower indices.
The terms in brackets define the curvature $R^A$ while the other two terms, using the antisymmetry of the structure constants in $(B,C)$ define the gauge covariant differential of $\epsilon^A$.
Therefore, using the \emph{anholonomized} parameter\footnote{By anholonomized parameter we mean that we are using the rigid group index of the vielbein $\mu^A $.} $\epsilon^A${,} the Lie derivative can be written as follows:
\begin{equation}\label{lie2}
{\ell}_{\epsilon} \mu^A= \left(\nabla \epsilon\right)^A + \iota_\epsilon R^A \,.
\end{equation}
Hence \textit{an infinitesimal diffeomorphism on the  manifold  $\rm\tilde G$ is a { $\rm G$}-gauge transformation plus curvature correction terms}.

In particular if the curvature $R^A$ has vanishing projection along the-  vector $\epsilon^{B}\tilde T_{ B}$, where $ B$ is an adjoint index of the subgroup $\rm H\subset \rm\tilde  G$ so that
\begin{equation}
\iota_\epsilon R^A \equiv \epsilon^{ B} R^A_{\phantom{A}{ B}C} \mu^C = 0,
 \end{equation}
then \emph{the action of the Lie derivative ${\ell}_{\epsilon}$ coincides with a gauge transformation}. In this case we recover the result that the  curvatures are \emph{horizontal} along the $H\subset G$ directions, in which case  the group manifold itself acquires the structure of a principal fiber bundle whose base manifold, (super)space, can be identified with $\tilde{\mathrm{G}}/H$,  $H$ being the gauge group.

We stress  that the derivation of the formula in equation  (\ref{lie2}), makes no explicit reference to the specific group  $\rm\tilde G$. It holds for any group, including supergroups, as we can see in the supergravity case.

\section{On the equivalence of FDA with the classical definition of L-infinity algebra.} \label{Linfinity}

We report in this Appendix part of the content of reference \cite{nlab}
 showing that the definition of CIS/FDA structures for the extension of Lie algebras to higher p-forms structures  \footnote{Actually, they are extended Chevalley-Eilenberg Lie algebras,
\commas {{{C}}hevalley-{{E}}ilenberg algebra}".} gives a dual formulation of an $L_\infty$-algebra.

To show this, let us first shortly remind the definition of an $L_\infty$ algebra.

An $L_{\infty}$ algebra is defined as:

\begin{itemize}
    \item a $\mathbb{Z}$-graded vector space ${g}$;
    \item For each $n\,\in\,\mathbb{N}$, a multilinear map $l_n$, called the \emph{n-ary bracket}, of the form

\begin{equation}
l_n(\dots)=[-,-,\dots,-]: g\bigotimes\dots \bigotimes g\rightarrow g
\end{equation}
\end{itemize}
and of degree $n-1$, such that the following conditions hold:

\begin{enumerate}
\item
({\bf {graded skew symmetry}}) :\\ each $l_n$ is graded antisymmetric, in that for every permutation $\sigma$
and for every n-tuple of homogenously graded $v_i \in\, g$ then:

\begin{equation}
l_n\left(v_{\sigma_1},v_{\sigma_2}\dots v_{\sigma_n}\right)=\chi(\sigma,v_1,\dots v_n)l_n \left(v_{1},v_{_2}\dots v_{_n}\right)
\end{equation}
where the graded signature $\chi(\sigma,v_1,\dots v_n)$ is defined as the product of the signature of the permutation times  a factor $(-1)^{|v^i| |v^j|}$ for each interchange of neighbours $\left(\dots v_i v_j\dots \right)$ to  $\left(\dots v_j v_i\dots \right)$ involved in the decomposition of the permutation as a sequence of swapping neighboring pairs. Note that this definition of $\chi$ matches our law sign of equation  \eqref{exchange}namely the Koszul sign law.

\item ({\bf {strong homotopy Jacobi identity}}):\\
For all $n \in \mathbb{N}$ and for all $n$-tuples $\left(v_{1},v_{_2}\dots v_{_n}\right)$ of homogeneously graded elements $v_i$, the following equation  holds:

\begin{equation}
\sum_{i,j (i+j=n+1)}\left[\sum_{\sigma\,\varepsilon Unsh(i,j-1)}\chi(\sigma,v_1,\dots v_n)\,(-1)^{i(j-1)}\,l_j\left(l_i\left(v_{\sigma_1},v_{\sigma_2}\dots v_{\sigma_i}\right),v_{\sigma_{i+1}},\dots v_{\sigma_n}\right)\right]=0 \label{strong}
\end{equation}
\end{enumerate}
where the sum \commas Unsh"= \commas Unshuffled" means  that we must sum over all the permutations of ($1,2,\dots,n$) that keep $i_1,\dots,i_{j}$ and $i_{j+1},\dots,i_{n}$ in the same relative order.\\

Actually, one equivalent definition can be obtained  passing to the degreewise finite dimensional \emph{dual graded vector space} of the Grassmann algebra of the $p$-forms, that is to the dual FDA. In this case one has a \emph{semifree differential graded algebra} and the Grassmann algebra is naturally equipped with an ordinary differential, namely the exterior derivative $d:d^2=0$.

The action of the differential $d$ on a set of basic elements $t^a$ of the Grassman algebra is written in the following way:
\begin{equation}
 dt^a=-\sum_{k=1}^\infty \frac{1}{k!} [t_{a_1}\dots t_{a_k}]^a\, t^{a_1}\wedge\dots \wedge t^{a_k}\,,
\end{equation}
where the multiple bracket $[t_{a_1}\dots t_{a_k}]^a$ is introduced.
Comparing with \eqref{deteta} we see that we can identify $ [t_{a_1}\dots t_{a_k}]^a$ with the generalized structure constants $C^{A(p)}_{B_{1(p_1)}B_{2(p_2)}\dots B_{n(p_n)}}$ of \eqref{deteta},
and that the $t^{a_i}$ span a basis of $p$-forms of  the Grassmann algebra, so that they can be identified with our    $\Theta^{B_{i}(p_i)}$  (the ${p_i}$ are the  form degree of $\Theta^{a_i}$).
The $d^2$ operator gives:
\begin{align}
&d\,d t^a= -d\sum_{k=1}^\infty \frac{1}{k!} [t_{a_1}\dots t_{a_k}]^a\, t^{a_1}\wedge\dots \wedge t^{a_k}= \nonumber \\
&^=\sum_{k,l}^\infty \frac{1}{k-1)!\,l!} [[t_{b_1}\dots t_{b_l}]\,t_{a_2}\dots t_{a_k}]^a\,t^{b_1}\wedge\dots\wedge t^{b_l}\wedge t^{a_2}\wedge\dots\wedge t^{a_k}=0\label{ddt}
\end{align}
which of course, given the previous identifications, coincides with \eqref{closure}.
 The important observation now is that the wedge products of the $t^{a_i}$ (as for the equivalent   $\Theta^{B_i(p_i)}$ forms) project the nested brackets onto their graded symmetric components. This occurs because one can sum over all permutations $\sigma$ of the $k+l-1$ indices weigthed with the Koszul phase of the permutation, which was identified, in the FDA formalism, with the phase ${(-1)^{B_i B_{i+1}+ p_i p_{i+1}}}= (-1)^\sigma$ as it is shown in \eqref{exchange}.

It follows that we can rewrite the right-hand side of \eqref{ddt} as follows:
\begin{align}
& \sum_{k,l}^\infty \frac{1}{(k+l-1)!}\sum_{\sigma \varepsilon Unsh(l,k-1)} (-1)^\sigma \frac{1}{(k-1)!\,l!}\,[[t_{b_1}\dots t_{b_l}]\,t_{a_2}\dots t_{a_k}]^a\wedge\dots\wedge t_{b_l}\wedge t^{a_2}\wedge\dots t^{a_k}=0\,.\label{ddt1}
\end{align}
Now the sum over all permutations can be decomposed into a sum over over the $(l,k-1)$ \emph{unshuffled},
and a sum over permutations inside the first $l$ and the last $k-1$ indices.
These latter permutations do not change the graded symmetry of the nested brackets, since the same permutation acts on the $t^{a_i}$ forms. As there are $(k-1)!\,l!$ of them, equation  \eqref{ddt1}
can be rewritten as follows:

\begin{align}
& \sum_{k,l=1}^\infty \frac{1}{(k+l-1)!}\sum_{\sigma \varepsilon Unsh \,(l,k-1)} (-1)^\sigma \,[[t_{a_1}\dots t_{a_l}]\,t_{a_{l+1}}\dots
t_{a_{k+l-1}}]
t^{a_1}\wedge\dots\wedge t^{a_{k+l-1}}=0\,.\label{ddt2}
\end{align}
Therefore the condition $d^2=0$ is equivalent to the condition
\begin{align}
\sum_{k+l=n-1}\sum_{\sigma \varepsilon Unsh\,(l,k-1)} (-1)^\sigma \,[[t_{a_1}\dots t_{a_l}]\,t_{a_{l+1}}\dots
t_{a_{k+l-1}}] =0
\end{align}
that reproduces the condition of strong homotopy identity \eqref{strong}, and therefore defines  an $L_{\infty}$ algebra.

\section{Coefficients in the Hidden-Algebra description of D=11 FDA}\label{coeff11}
To satify equation \eqref{da3par}, the coefficients $T_i,S_j,E_j$, with $i=1,\cdots ,5$, $j=1,2,3$, should satisfy the following set of algebraic equations (from \cite{Andrianopoli:2016osu},\cite{Andrianopoli:2017itj}):
\begin{equation} \label{cond11}
\left\{
\begin{array}{l}
T_0-2 S_1 E_1-1=0\,,\cr
T_0-2 S_1 E_2 -2 S_2 E_1=0\,,\cr
3 T_1-8 S_2 E_2=0\,,\cr
T_2+10 S_2 E_3+10 S_3 E_2=0\,,\cr
120 T_3-S_3 E_1-S_1 E_3=0\,,\cr
T_2+1200 S_3 E_3 =0\,, \cr
T_3-2S_3 E_3=0\,,\cr
9T_4+10 S_3 E_3=0\,,\cr
S_1+10 S_2-720 S_3=0\,,\cr E_1+10 E_2-720 E_3 = 0\,,
\end{array}
\right.
\end{equation}
which are solved by:
\begin{align}
&\left\{\begin{array}{ccl}
T_0 &=& \frac{1}{6}+ \alpha, \\
T_1 &=&-\frac{1}{90} + \frac{1}{3} \alpha, \\
T_2 &=&-\frac 1{4!} \alpha, \\
T_3 &=& \frac{1}{(5!)^2}\alpha , \\
T_4 &=& - \frac{1}{3[2!\cdot (3!)^2 \cdot 5!]} \alpha ,
\end{array}\right.
\\
&
\left\{\begin{array}{ccl}
S_1 &=&
\frac 1{2C}\left(\frac{10}{5!} +\sqrt{\frac\alpha{5!}}\right)\,,\\
S_2 &=& \frac 1{2C}\left(-\frac{1}{5!} +\frac 12\sqrt{\frac\alpha{5!}}\right)\,,\\
S_3 &=& \frac 1{2C}\frac{1}{5!}\sqrt{\frac\alpha{5!}}
\end{array}\right.
\label{S11}\\
&
\left\{\begin{array}{ccl}
E_1 &=& 5!  C \left(-\frac{10}{5!} +\sqrt{\frac\alpha{5!}}\right)\,,\\
E_2 &=& 5!  C \left(\frac{1}{5!} +\frac 12\sqrt{\frac\alpha{5!}}\right)\,,\\
E_3 &=& 5! C\,\left(\frac 1{5!}\sqrt{\frac\alpha{5!}}\right)\,,
\end{array}\right.\label{E11}
\end{align}
 $\alpha$ being a free parameter on which the hidden algebra discussed in Section \ref{hidden} depends, while $C$ is a spurious coefficient due to the fact that equations \eqref{cond11} contain the parameters $S_i$ and $E_j$ (with $i,j=1,2,3$) always homogeneously, or  in the combination $S_iE_j$. For the same reason, given the solutions \eqref{S11} and \eqref{E11}, also the set of coefficients with interchanged values $2C\,S_i \leftrightarrow -\frac 1{5!C}\,E_i$ is an equivalent solution to \eqref{cond11}. A particularly symmetric choice of $C$ is $C=i\frac{1}{\sqrt{5!2!}}$.
 Finally, we note that  the relations presented here, and in particular \eqref{S11} and \eqref{E11}, look different, and are in fact more explicit,  from the equivalent formulas in \cite{Andrianopoli:2017itj}.

$$
$$

\end{document}